\documentclass[9pt,twocolumn,twoside]{opticajnl}
\journal{opticajournal} %

\setboolean{shortarticle}{false}

\usepackage{lineno}

\usepackage{bm}
\usepackage{mathabx}
\usepackage{tikz}
\usepackage{multirow}
\usepackage[export]{adjustbox}
\usetikzlibrary{calc, positioning, arrows.meta, angles, quotes}

\usepackage{titlesec}
\usepackage{hyperref}
\makeatletter
\renewcommand{\p@subsection}{\thesection.} %
\makeatother

\newcommand{\networkname}{kFinder}

\title{Inverse Synthetic Aperture Fourier Ptychography}

\author[1,*]{Matthew~A.~Chan}
\author[2]{Casey~J.~Pellizzari}
\author[1]{Christopher~A.~Metzler}

\affil[1]{Department of Computer Science, University of Maryland, College Park, MD, 20742}
\affil[2]{United States Air Force Academy, Air Force Academy, CO, 80840}

\affil[*]{mattchan@umd.edu}

\begin{abstract}
Fourier ptychography (FP) is a powerful light-based synthetic aperture imaging technique that allows one to reconstruct a high-resolution, wide field-of-view image by computationally integrating a diverse collection of low-resolution, far-field measurements. 
Typically, FP measurement diversity is introduced by changing the angle of the illumination or the position of the camera; either approach results in sampling different portions of the target's spatial frequency content, but both approaches introduce substantial costs and complexity to the acquisition process. 
In this work, we introduce Inverse Synthetic Aperture Fourier Ptychography, a novel approach to FP that foregoes changing the illumination angle or camera position and instead generates measurement diversity through target motion. 
Critically, we also introduce a novel learning-based method for estimating k-space coordinates from dual plane intensity measurements, thereby enabling synthetic aperture imaging {\em without knowing the rotation of the target}. 
We experimentally validate our method in simulation and on a tabletop optical system.%
\end{abstract}

\setboolean{displaycopyright}{false} %

\begin{document}

\maketitle

\begin{figure*}[t]
    \centering
    \begin{tikzpicture}
        \coordinate (O) at (0,0);
        \coordinate (A) at (0,0.5);
        
        \node[] at (-2.5, 4.5) {Rotating Target};
        \draw[thick, blue] (0,0) -- (0,4) node[above] {Angle 1};
        \draw[thick, orange, rotate around={30:(0,2)}] (0,0) -- (0,4) node[above, left] {Angle 2};
        \draw[thick, purple, rotate around={60:(0,2)}] (0,0) -- (0,4) node[above, left] {Angle 3};
        \draw[black] (0,1.5) arc (-90:-90+80:0.2) node[midway, below] {$\theta$};

        \fill[green] (-3, 0.5) circle (2pt);
        \node[left] at (-3.2, 0.52) {Laser};
        \draw[green] (-2.85,0.5) -- (-2.75,0.5);
        \draw[green, rotate around={30:(-3, 0.5)}] (-2.85,0.5) -- (-2.75,0.5);
        \draw[green, rotate around={-30:(-3, 0.5)}] (-2.85,0.5) -- (-2.75,0.5);
        \draw[green, dashed, ->] ($(A) + (-2.5,0)$) -- ($(A) + (0.8,0)$);

        \draw[black, |-|] (-0.5,0.6) -- (-0.5,2) node[midway,left] {$s$};
        \draw[black, |-|] ($(A) + (0, -0.8)$) -- ($(A) + (0.8, -0.8)$)node[midway, below,yshift=-4] {$s\tan\theta$};

        \node[left] at (4.5, 3.7) {Fourier Plane};
        \node[left] at (4.5, 0.7) {Image Plane};
        \node[above] at (5.9, 4.7) {\textcolor{blue}{Angle 1}};
        \node[above] at (8.6, 4.7) {\textcolor{orange}{Angle 2}};
        \node[above] at (11.3, 4.7) {\textcolor{purple}{Angle 3}};

        \node[inner sep=0pt] () at (8.5,2){\includegraphics[width=.45\textwidth]{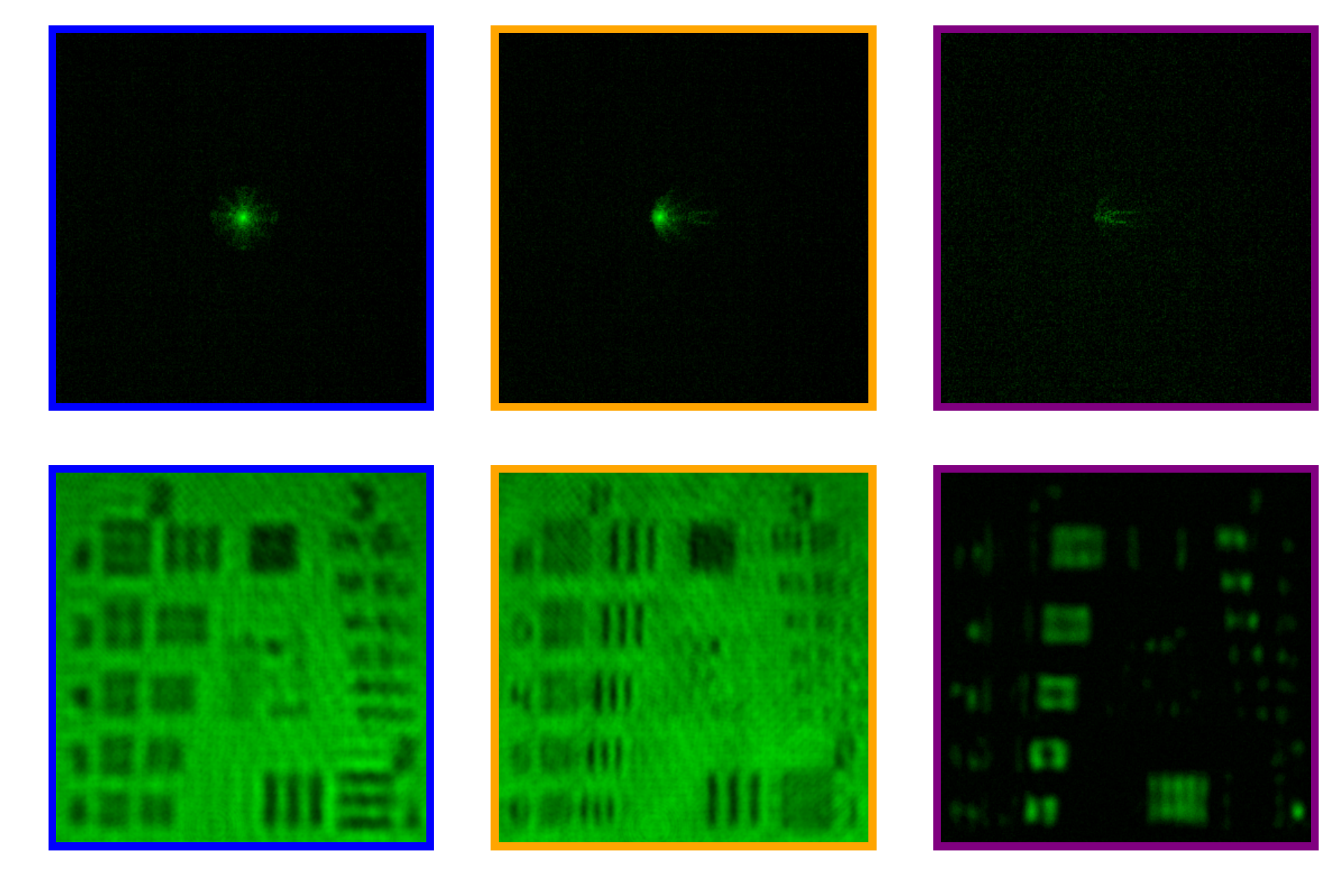}};
    \end{tikzpicture}
    \caption{\textbf{Rotation-induced phase shift.} Rotating a reflective target increases the optical path length, imparting a linear phase shift. At the Fourier plane, this equates to a translation of the spectrum. By rotating across various angles, we effectively scan different areas of the target's spectrum.}
    \label{fig:rotation_illustration}
\end{figure*}
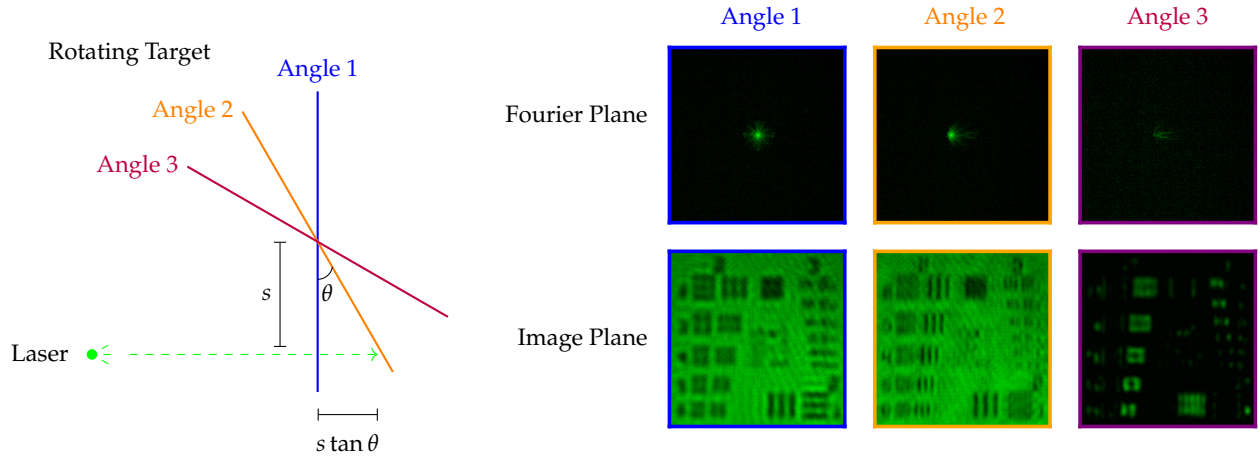

\section{Introduction}
Optical imaging systems, from microscopes to telescopes, have long been cornerstones of scientific discovery, enabling researchers to observe and study structures across vastly different scales. However, these systems face fundamental physical limitations that constrain image quality. Their resolution is determined by their numerical aperture (NA), which defines the range of angles over which the system can accept or emit light~\cite{goodmanIntroductionFourierOptics1969,ouHighNumericalAperture2015}. While high NA objectives offer superior resolution, they typically suffer from a limited field-of-view (FoV)---forcing a trade-off between resolution and the area that can be imaged at once.

Fourier ptychography (FP)~\cite{zhengWidefieldHighresolutionFourier2013,zhengConceptImplementationsApplications2021} is an innovative light-based synthetic aperture imaging method which overcomes this tradeoff. 
By changing the angle at which a sample is illuminated or by translating the objective lens, FP records a series of measurements, each of which captures distinct information about the sample's spatial frequency content. Phase retrieval algorithms are then used to combine these measurements and form a single high-resolution, wide-FoV image of the sample. %

While FP has seen widespread success in microscopy, it has seen limited uptake in macroscopic applications such as astronomy. This is, in part, because FP requires either translating the objective---which is impractical with large telescopes---or controlling the illumination angle of the target---which is infeasible for distant celestial objects like satellites.

In this work we introduce a novel inverse synthetic aperture (ISA) imaging~\cite{chen1980target} technique which leverages angular diversity caused by target motion (i.e., rotation) rather than the illumination to perform FP. Our approach eliminates the need for direct control of the illumination source, enabling high-resolution, wide-FoV imaging in situations where illumination is difficult to manipulate. Our method is especially well-suited to various astronomical imaging applications where orbital motion naturally introduces measurement diversity. However, this application presents a unique challenge because the motion-induced spatial frequency shifts are generally unknown---that is, one does not control or even know the target's trajectory.%

To address this challenge, we propose \networkname{}, a novel neural network (NN) that estimates spatial frequency shifts from dual plane amplitude measurements. This network serves as a learned k-space localization module for FP and demonstrates strong generalization capabilities across the domain of natural images. As a result, our method is effective in opportunistic imaging scenarios where the target's rotation is unknown. Note our approach could also be applied in conventional FP settings, where it could relax calibration requirements on the illumination source.

In this work, we make the following contributions:
\begin{itemize}
        \item We introduce Inverse Synthetic Aperture Fourier Ptychography (ISA-FP), a novel FP technique that introduces angular diversity via target rotation. 
        \item We present \networkname{}, a novel neural network that estimates k-space aperture locations from dual plane amplitude images. This unlocks calibration-free FP and suggests that natural images contain learnable structure in their spectra.
        \item We validate our approach in simulation and show that our calibration-free method achieves super-resolution performance on-par with calibrated methods.
        \item We construct a dual plane imaging system on an optical workbench and demonstrate $3.6\times$ super-resolution of a resolution target.
        \item We characterize the importance of various image features on \networkname{}'s convergence and find that pupil plane features are crucial for convergence.
\end{itemize}

\section{Background: Fourier Ptychography}

\subsection{Measurement Process}
FP captures a diverse collection of passband filtered measurements, each recording different information about the sample's frequency content. Traditional FP systems achieve measurement diversity by controlling the angle $\bm{\theta} = [\theta_x, \theta_y]^T$ at which a sample is illuminated. 
Mathematically, when a thin sample $o(\bm{s})$ with spatial coordinates $\bm{s} = [s_x, s_y]^T$ is illuminated by a plane wave with an angle $\bm{\theta}_j$ for $j = 1, 2, \cdots, N$, it is modulated by a linear phase ramp
\begin{equation}
  \exp\left(i\bm{s} \cdot \frac{2\pi}{\lambda}\sin \bm{\theta}_j \right),
\end{equation}
where $\lambda$ is the wavelength of light~\cite{goodmanIntroductionFourierOptics1969}. 

At the pupil (or Fourier) plane, this shifts the sample's spectrum by a wavevector $\bm{k}_j = \frac{2\pi}{\lambda}\sin \bm{\theta}_j$, such that one records the translated signal
\begin{equation}
  O(\bm{k} - \bm{k}_j) = \mathcal{F}\left\{o(\bm{s})\exp\left(i \bm{k}_j \bm{s} \right)\right\},
  \label{eq:pupil_plane}
\end{equation}
where $O(\bm{k})=\mathcal{F}\{o(\bm{s})\}$ and $\bm{k} = [k_x, k_y]^T$ are frequency coordinates~\cite{voelzComputationalFourierOptics2011,schmidtNumericalSimulationOptical2010}.
This translated field then passes through a lens and one records an intensity measurement
\begin{equation}
    I_j(\bm{s}) = \left|\mathcal{F}^{-1}\left\{O(\bm{k} - \bm{k}_j)M(\bm{k})\right\}\right|^2,
    \label{eq:image_plane}
\end{equation}
where $M(\bm{k})$ is the system's pupil function (typically a circular mask). By varying the illumination angle $\bm{\theta}_j$, we can effectively scan different regions of $O(\bm{k})$ through the fixed pupil $M(\bm{k})$.
Alternatively, one could translate the aperture to location $\bm{k}_j$ in k-space. The two approaches, tilting the illumination or translating the aperture, are mathematically equivalent. 

\subsection{Phase Retrieval}
With a sequence of intensity-only measurements $I_j(\bm{s})$ in hand, the goal of FP is to computationally recover the high-resolution complex field $o$, or equivalently its Fourier transform $O$. This is accomplished by solving a ``phase retrieval'' problem. For example one can minimize
\begin{equation}
  \underset{\hat{O}}{\arg\min} \sum_{j} \left\|I_j(\bm{s}) - \left|\mathcal{F}^{-1}\left\{\hat{O}(\bm{k} - \bm{k}_j)M(\bm{k})\right\}\right|^2\right\|_2^2,
  \label{eq:fp_objective}
\end{equation}
or similar loss functions using any number of algorithms~\cite{yehExperimentalRobustnessFourier2015}. 
These algorithms assume precise, or at least general~\cite{eckertEfficientIlluminationAngle2018}, knowledge of the wavevectors/frequency-shifts $\bm{k}_j$ associated with each measurement. 
Our work seeks to eliminate this assumption.

\begin{figure*}[!t]
    \centering
    \begin{tikzpicture}
        \node[] at (-6.2, 8.2) {(a)};
        \node[] at (3.3, 8.2) {(b)};
        
        \node[inner sep=0pt] () at (-6.0,3) {\includegraphics[width=.4\textwidth]{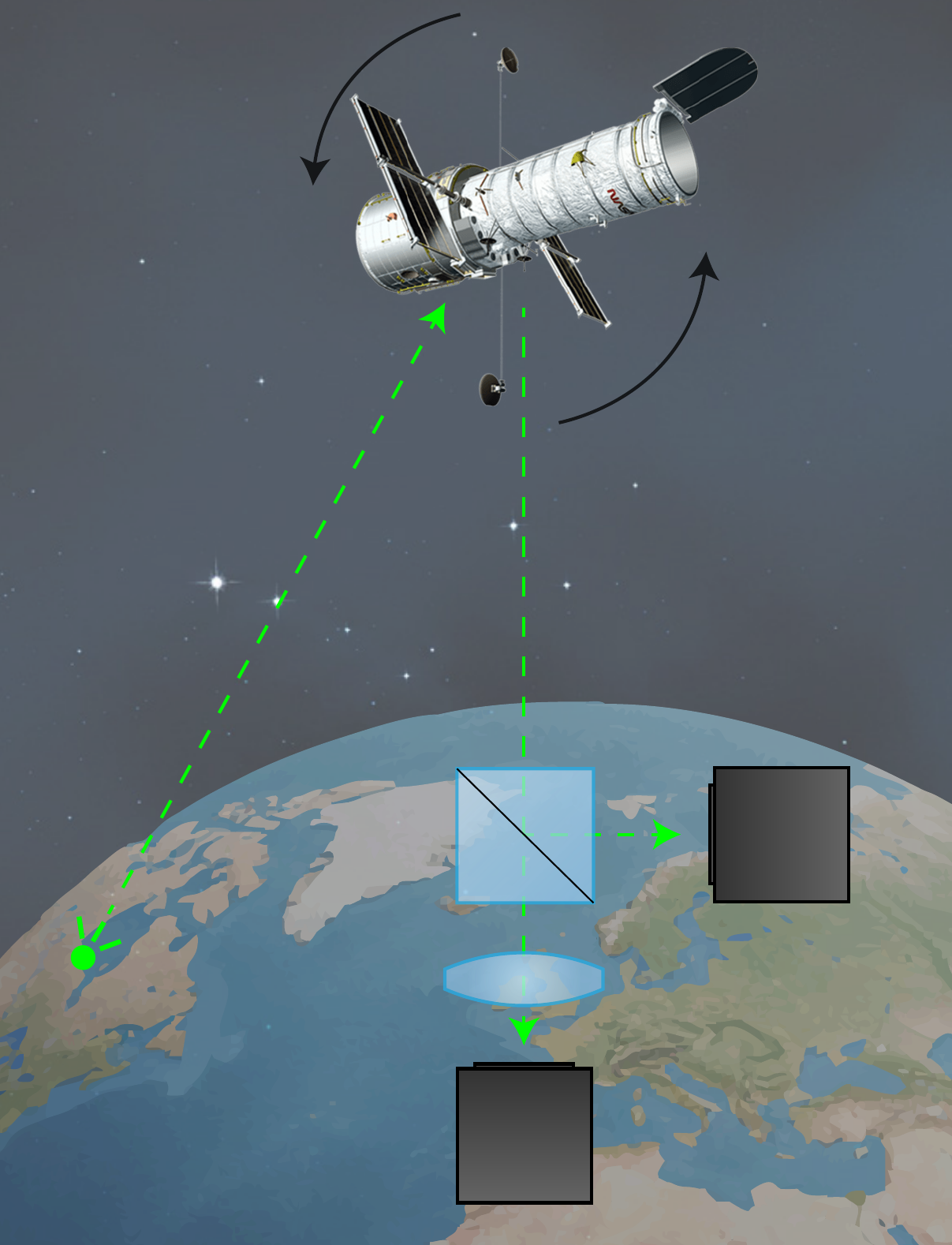}};
        \node[] at (-8.2, 5.8) {\bf\textcolor{white}{Rotating Target}};
        \node[] at (-8.5, 1.0) {\bf\textcolor{white}{Laser}};
        \node[text width=1cm, align=center] at (-6.6, 1.3) {\bf\textcolor{white}{Beam \\ Splitter}};
        \node[align=center] at (-6.6, 0.2) {\bf\textcolor{white}{Lens}};
        \node[text width=1cm, align=center] at (-6.6, -1.0) {\bf\textcolor{white}{Image \\ Camera}};
        \node[text width=1cm, align=center] at (-3.3, 0.2) {\bf\textcolor{white}{Pupil \\ Camera}};

        \node[rotate=90] at (-1.8, 6.0) {Dual Plane Images};
        \node[inner sep=0pt] () at (-0.5,7) {\includegraphics[width=.1\textwidth]{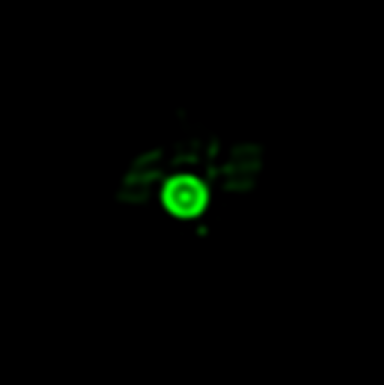}};
        \node[inner sep=0pt] (nn_input) at (-0.5,5) {\includegraphics[width=.1\textwidth]{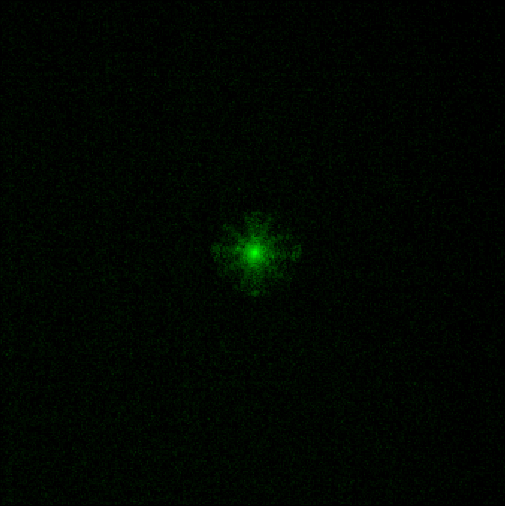}};
        \draw[thick, black, ->] (0.5,6.0) -- (0.7,6.0) node[] {};

        \node[] at (3.0, 7.1) {\networkname{}};
        \node[inner sep=0pt] () at (3.0,5.93) {\includegraphics[width=.25\textwidth]{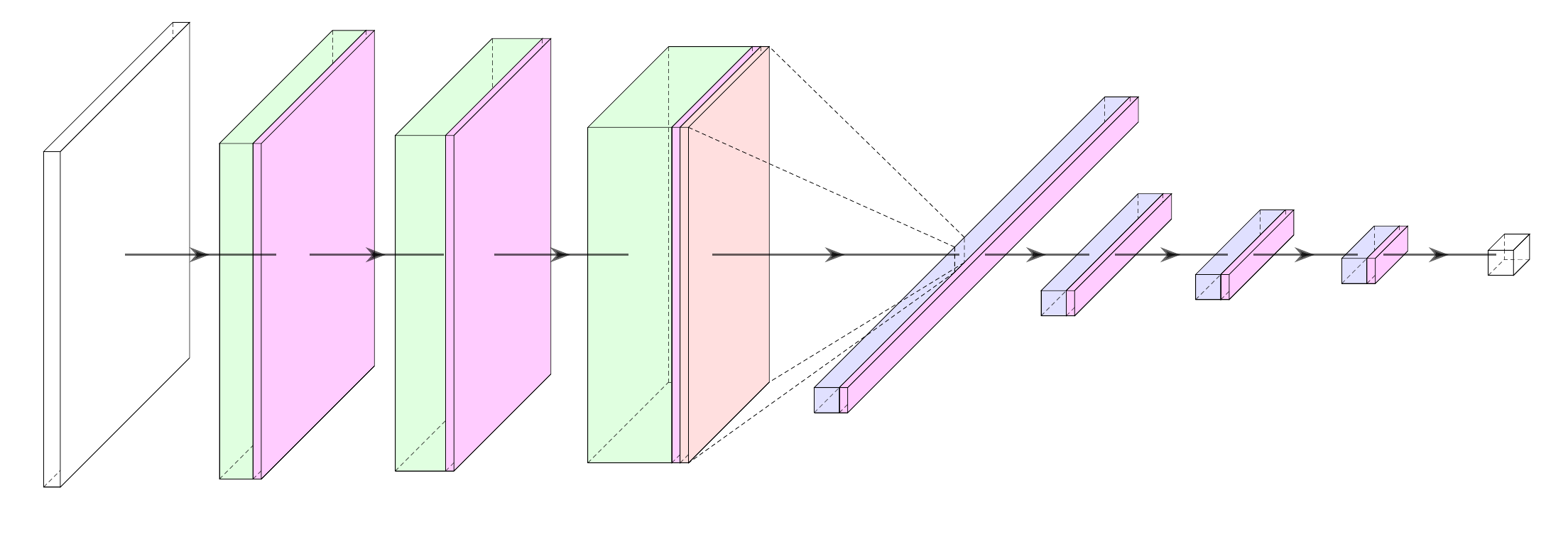}};

        \draw[thick, black, ->] (5.3,6.0) -- (5.5,6.0) node[] {};
        
        \node[] at (7.0, 7.8) {Estimated k-Space};
        \node[inner sep=0pt] (nn_output) at (7.0,6) {\includegraphics[width=.16\textwidth]{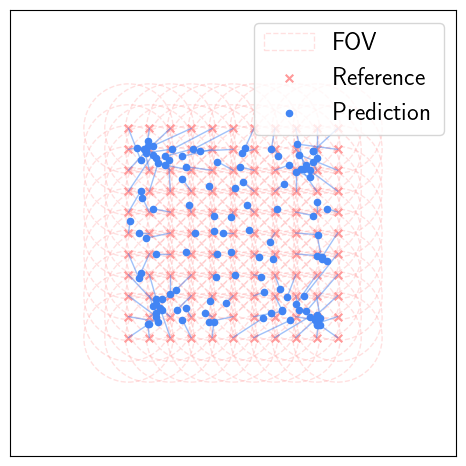}};

        \node[] at (3, 3.9) {Iterative Solver};
        \node[draw] (solver) at (3, 3.0) {$\underset{\hat{O}, \hat{\bm{k}}_j}{\arg\min} \sum_{j} \left\|I_j(\bm{s}) - \left|\mathcal{F}^{-1}\left\{\hat{O}(\bm{k} - \hat{\bm{k}}_j)M(\bm{k})\right\}\right|^2\right\|_2^2$};

        \node[] at (-0.2, 1.8) {Recovered Amplitude\strut};
        \node[] at (3.3, 1.8) {Recovered Phase\strut};
        \node[] at (6.8, 1.8) {Corrected k-Space\strut};
        \node[inner sep=0pt] () at (-0.2,-0.2) {\includegraphics[width=.18\textwidth]{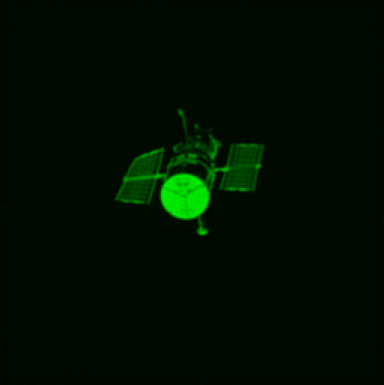}};
        \node[inner sep=0pt] () at (3.3,-0.2) {\includegraphics[width=.18\textwidth]{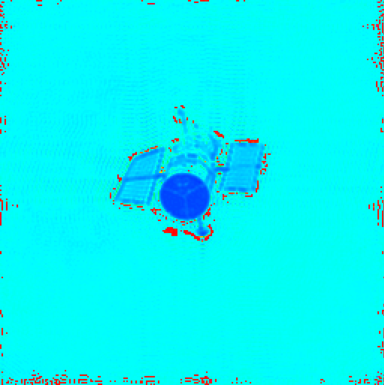}};
        \node[inner sep=0pt] () at (6.8,-0.2) {\includegraphics[width=.19\textwidth]{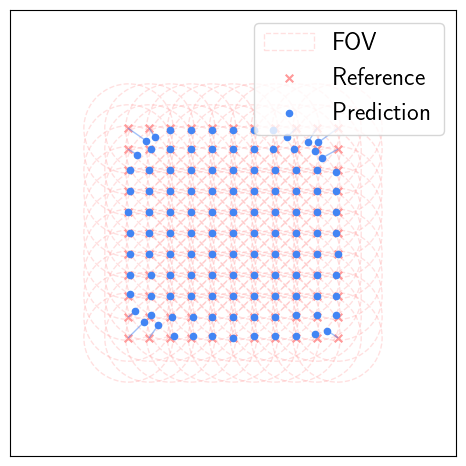}};

        \draw[thick,->] (nn_input.250) |- (solver.west);
        \draw[thick,->] (nn_output.280) |- (solver.east);
    \end{tikzpicture}
      \caption{\textbf{Inverse Synthetic Aperture Fourier Ptychography.} (a) A rotating target is illuminated and the resulting wavefront is imaged at both pupil and image planes across various rotation angles. (b) Dual plane intensity measurements are fed into a neural network that estimates corresponding k-space locations. An iterative solver jointly optimizes both the wavefront and k-space locations.}
      \label{fig:main_fig}
\end{figure*}
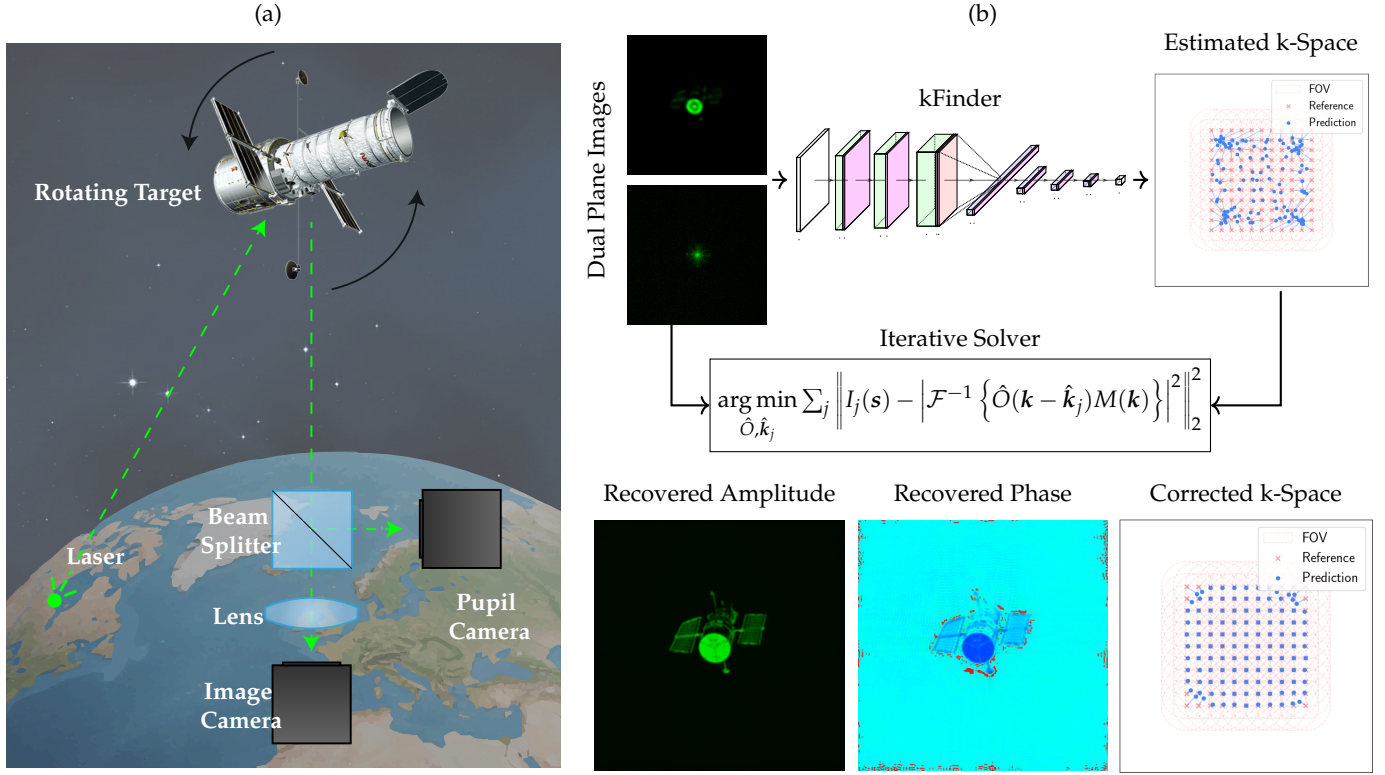

\section{Proposed Method}

ISA-FP differs from conventional FP in two key respects. 
First, measurement diversity is achieved not through tilting the illumination or translating the aperture, but through rotating the sample. 
Second, we no longer assume that the frequency shifts associated with each measurement are known.

\subsection{Measurement Diversity Through Rotation}

Rather than modeling frequency shifts caused by the source illumination angle $\bm{\theta}$ or translating the aperture location, we instead consider frequency shifts caused by the rotation $\bm{\theta}$ of a reflective sample illuminated by plane waves parallel to the optical axis (see Figure~\ref{fig:rotation_illustration}). This rotation introduces a difference in optical path length of incident light at different positions $\bm{s}$ on the sample, resulting in a sample-plane phase shift of
\begin{equation}
  \exp\left(i \bm{s} \cdot \frac{4\pi}{\lambda}\tan\bm{\theta}\right)
  \label{eq:phase_shift_sample_tilt}.
\end{equation}
Note that the extra factor of 2 arises because the sample is reflective---both the incident and reflected waves experience extra path length due to the tilted surface.

For small angles $\theta \ll 1$, we can apply the approximation $\tan\theta \approx \theta$, yielding a simplified phase shift of
\begin{equation}
   \exp\left(i \bm{s} \cdot \frac{4\pi}{\lambda}\bm{\theta}\right).
  \label{eq:phase_shift_sample_tilt_approx}
\end{equation}
Empirically, we find that this approximation is well-justified, as angles $\theta < 0.01$ radians provide multiple apertures worth of coverage of the sample's Fourier spectrum across a wide variety of optical configurations. Importantly, at these angles, geometric distortions at the image plane, such as stretching and shrinking, are negligible. 
As such, our measurements follow the same model (\eqref{eq:image_plane}) as in conventional FP, but with the spatial frequency shift, $\bm{k}_j = \frac{4\pi}{\lambda}\bm{\theta}_j$, now a function of the {\em unknown} target rotation.

\begin{figure*}
    \centering
    \includegraphics[width=0.9\textwidth]{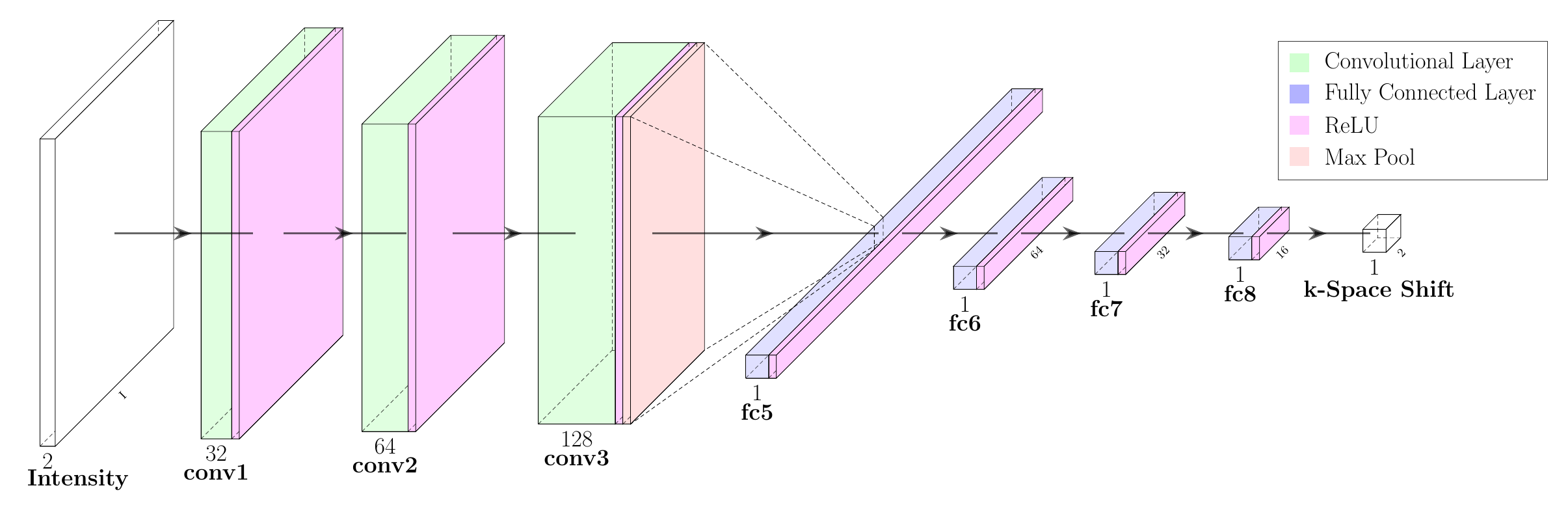}
    \caption{\textbf{\networkname{} architecture.} Our k-space localization module, \networkname{}, is a series of convolutional and fully connected layers that process dual plane intensity measurements and predict 2D frequency-shifts.}
    \label{fig:network_architecture}
\end{figure*}

\subsection{FP with Unknown Spatial Frequency Shifts}

In this work we simultaneously reconstruct the sample wavefront $\hat{O}$ and the spatial frequency shifts $\hat{\bm{k}}_j$ by minimizing the objective
\begin{equation}
  \mathcal{L}_\text{data}(\hat{O}, \hat{\bm{k}}_1,...\hat{\bm{k}}_N) = \sum_{j} \left\|I_j(\bm{s}) - \left|\mathcal{F}^{-1}\left\{\hat{O}(\bm{k} - \hat{\bm{k}}_j)M(\bm{k})\right\}\right|^2\right\|_2^2.
  \label{eq:data_objective}
\end{equation}

We also include two additional regularization terms to the objective: (\textit{i}) a total variation (TV) loss~\cite{rudinNonlinearTotalVariation1992} to encourage smoothness in the target amplitude and (\textit{ii}) a sparsity loss on the target's phase to encourage zero phase in regions with zero intensity (which are otherwise unconstrained). These regularization terms are given by
\begin{align}
  \mathcal{L}_\text{TV}(\hat{O}) &= \left\|\nabla \hat{o}(\bm{s})\right\|_1, \text{ and }\\
  \mathcal{L}_\text{phase}(\hat{O}) &= \left\|\angle(\hat{o}(\bm{s}))\right\|_2^2,
\end{align}
where $\nabla(\cdot)$ denotes the spatial gradient operator and $\angle(\cdot)$ denotes the phase of a complex number.

Our combined objective is thus given by
\begin{equation}
  \underset{\hat{O}, \hat{\bm{k}}_1,...\hat{\bm{k}}_N}{\arg\min} \quad \alpha \mathcal{L}_\text{data}(\hat{O}, \hat{\bm{k}}_1,...\hat{\bm{k}}_N) + \beta \mathcal{L}_\text{TV}(\hat{O}) + \gamma \mathcal{L}_\text{phase}(\hat{O})
  \label{eq:our_objective}
\end{equation}
for weights $\alpha, \beta, \gamma \in [0,1]$.

\subsection{\networkname}

The ISA-FP objective function in~\eqref{eq:our_objective} is extremely non-convex. 
Consequently, minimizing it directly, without careful initialization, is ineffective. 
To overcome this, we train a neural network to perform initial k-space localization. Once trained, this network can be used to generate initial estimates of $\hat{\bm{k}}_j$.%

\subsubsection{Input Features}
\label{sec:input_features}

Prior work has demonstrated that pupil plane features contain useful information for estimating frequency shifts. In particular, Eckert et al.~\cite{eckertEfficientIlluminationAngle2018} demonstrated that pupil plane autocorrelation
\begin{equation}
  R_j(\bm{k}) = \mathcal{F}\left\{I_j(s)\right\}
  \label{eq:autocorrelation}
\end{equation}
contains structure within the bright-field region that can be used to calibrate $\bm{k}_j$. 

Inspired by this, we train a network---which we refer to as \networkname{}---to initialize wavevectors $\hat{\bm{k}}_j$ from pupil plane intensity measurements
\begin{equation}
  P_j(\bm{k}) = \left|O(\bm{k})M(\bm{k}-\bm{k}_j)\right|^2.
\end{equation} 
In particular, we provide the network with normalized pupil plane intensity $P_j(\bm{k})/\max(P_j(\bm{k}))$ and 
image plane intensity $I_j(\bm{s})$ as input. The normalization process forces the network to learn structural priors instead of relying on amplitude information, and we empirically find this to improve \networkname{}'s generalization capabilities.

\subsubsection{Network Architecture}

\networkname{} processes dual-channel (i.e., image plane and pupil plane) images through a four-layer convolutional neural network (CNN)~\cite{lecunGradientbasedLearningApplied1998,NIPS2012_c399862d} with increasing channel depths (2, 32, 64, 128) and ReLU activation functions~\cite{agarap2018deep}. The last convolutional layer has a max-pooling operation to reduce intermediate feature dimensions before they are flattened. Feature vectors are passed through four fully connected (FC) layers with output sizes (64, 32, 16, 2) to finally output 2D frequency-shifts $\hat{\bm{k}}_j \in \mathbb{R}^2$. In total, the network has approximately 134 million trainable parameters. See Figure~\ref{fig:network_architecture} for a detailed architecture diagram.

\begin{figure*}[t]
    \centering
    \begin{tikzpicture}
        \node[inner sep=0pt] () at (-0.6,0) {\includegraphics[width=.7\textwidth]{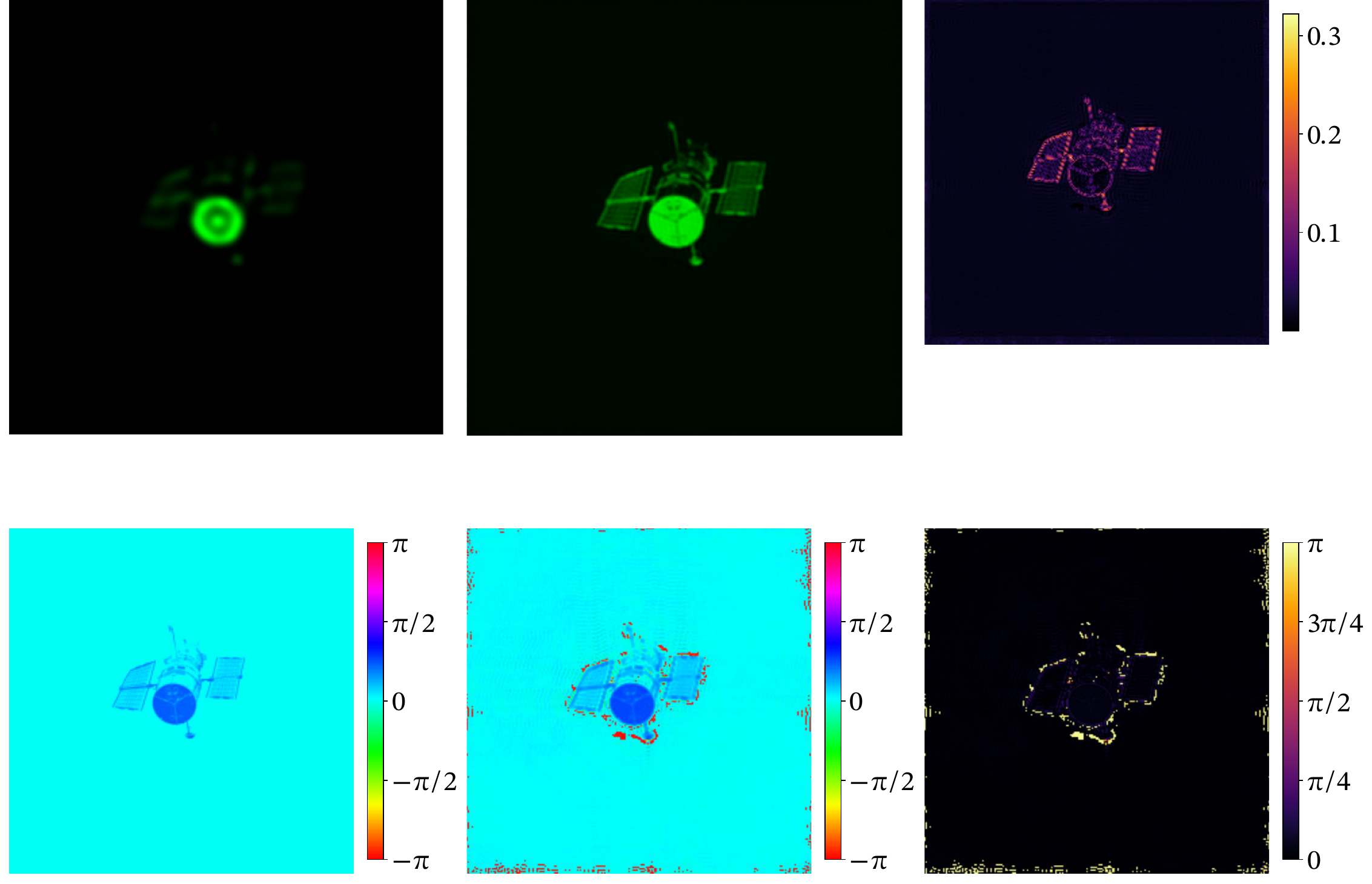}};
        \node[inner sep=0pt] () at (8.6,0) {\includegraphics[width=.21\textwidth]{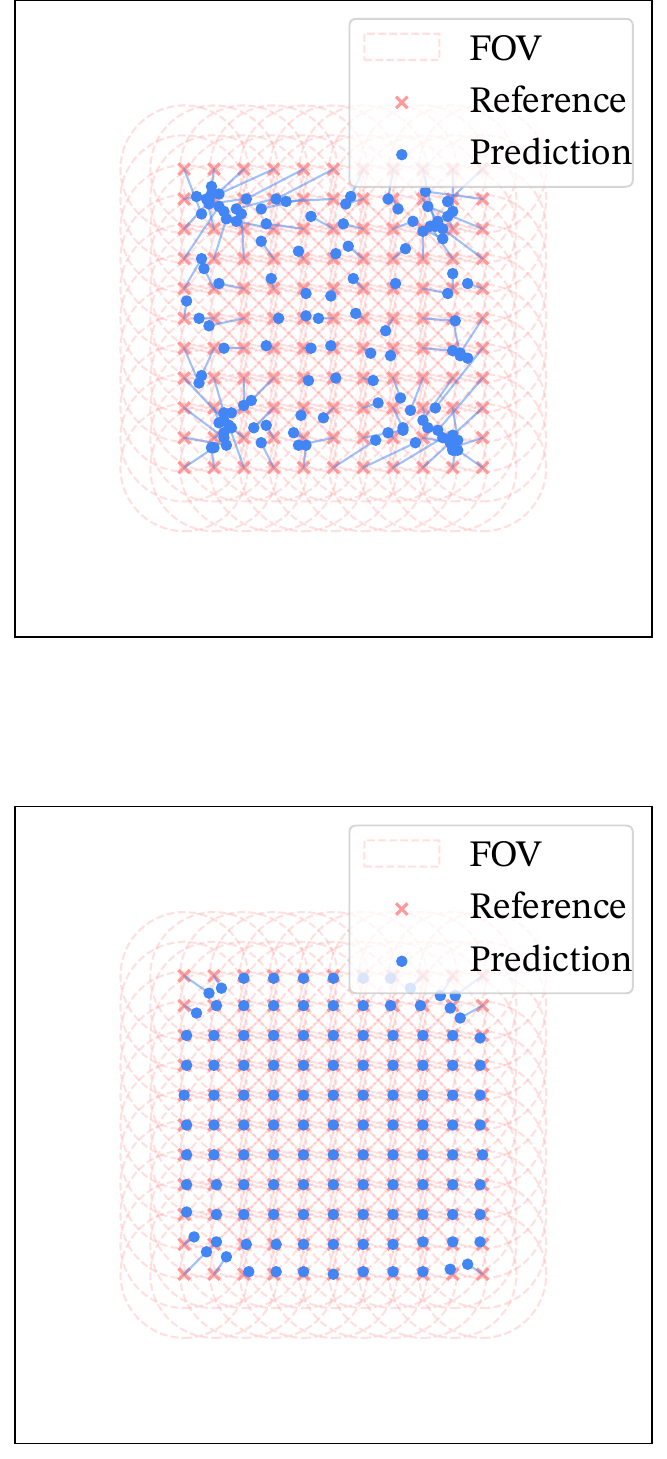}};
        \node[above] at (-0.5, 5) {(a)};
        \node[above] at (8.7, 5) {(b)};
        \draw[thick, white, |-|] (-2.3,0.5) -- (-1.3,0.5) node[midway,above] {25m};

        \node[above] at (-4.9, 4.2) {Low Resolution Image\strut};
        \node[above] at (-0.5, 4.2) {Super-resolved\strut};
        \node[above] at (3.3, 4.2) {Amplitude Error\strut};

        \node[above] at (-5.3, -0.7) {Ground-truth Phase\strut};
        \node[above] at (-1.0, -0.7) {Recovered Phase\strut};
        \node[above] at (3.4, -0.7) {Phase Error\strut};

        \node[above] at (8.7, 4.2) {\networkname{} Output};
        \node[above] at (8.7, -0.4) {Corrected k-Space};
    \end{tikzpicture}
    \caption{\textbf{Qualitative simulation results.} We simulate low resolution measurements using rotation angles sampled on a uniform grid. \networkname{} produces initial k-space estimates, which are corrected during optimization. Error plots show that our method accurately recovers the target's amplitude and phase.}
    \label{fig:sim_comparison}
\end{figure*}
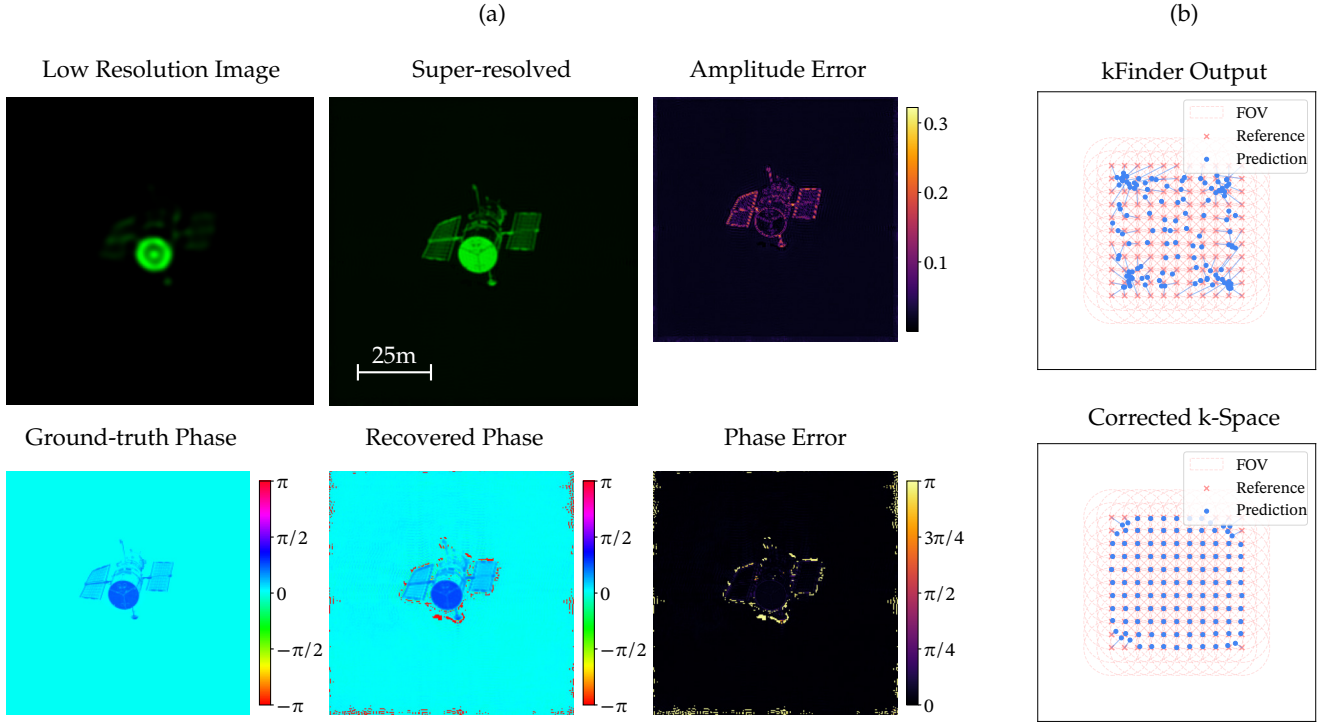

\subsection{Reconstruction Algorithm}
\label{sec:reconstruction}

After initializing $\hat{\bm{k}}_j$ using \networkname{}, we initialize $\hat{O}$ using any image plane measurement
$I_j(\bf{s})$ and reconstruct the sample's wavefront by alternating between three main steps:
\begin{enumerate}
  \item Project $\hat{O}(\bm{k})$ to the image plane and enforce an intensity constraint.
  \item Project the constrained estimate back to the pupil plane, and update $\hat{O}(\bm{k})$ via gradient descent on Equation~\eqref{eq:our_objective}.
  \item Search over a local window around $\hat{\bm{k}}_j$ to find the phase shift $\hat{\bm{k}}_j \pm \bm{\delta}$ which most closely matches the measured data.
\end{enumerate}

\subsubsection{Image Plane Projection}
We first project the estimated spectrum $\hat{O}$ to the image plane to obtain the wavefront
\begin{equation}
        \psi_j(\bm{s})=\mathcal{F}^{-1}\left\{\hat{O}(\bm{k} - \hat{\bm{k}}_j)M(\bm{k})\right\}.
        \label{eq:phase_image}
\end{equation}
Similar to Gerchberg-Saxton phase retrieval methods~\cite{jamesr.fienupPhaseRetrievalAlgorithms1982}, we apply an intensity constraint to the image plane wavefront, yielding
\begin{equation}
        \psi_j(\bm{s}) = \sqrt{\frac{I_j(\bm{s})}{\left|\psi_j(\bm{s})\right|^2}} \psi_j(\bm{s}).
        \label{eq:intensity_constraint}
\end{equation}
This effectively replaces the estimated amplitude with the true measured amplitude.

\subsubsection{Pupil Plane Projection}

We then project $\psi_j(\bm{s})$ back to the pupil plane to obtain
\begin{equation}
        \Psi(\bm{k}-\hat{\bm{k}}_j) = \mathcal{F}^{-1}\left\{\psi_j(\bm{s})\right\}.
        \label{eq:pupil_projection}
\end{equation}
Our method assumes access to dual plane intensity, so we can (optionally) enforce an additional intensity constraint 
\begin{equation}
        \Psi(\bm{k}-\hat{\bm{k}}_j) = \sqrt{\frac{P_j(\bm{k})}{\left|\Psi(\bm{k} - \hat{\bm{k}}_j)\right|^2}} \Psi(\bm{k} - \hat{\bm{k}}_j).
\end{equation}
at the pupil plane.
Although this constraint has been shown to improve convergence~\cite{fienup1978reconstruction}, we find that it overconstrains the degrees of freedom on the wavefront's amplitude. This causes any reconstruction errors to manifest in the phase component, rendering our sparse phase regularization ineffective. We omit this additional constraint in our experimental implementation.

After projection, we perform gradient descent on our objective function in Equation~\eqref{eq:our_objective} with respect to $\hat{O}$. Specifically, we compute the gradients of the data fidelity loss, total variation regularization loss, and phase sparsity loss with respect to $\hat{O}$. The second-order derivative of the data fidelity loss is given by
\begin{equation}
  \frac{\partial^2 \mathcal{L}_\text{data}}{\partial \hat{O}^2} =  \frac{\sum_j \widebar{M}(\bm{k}+\hat{\bm{k}}_j)}{\sum_j |M(\bm{k}+\hat{\bm{k}}_j)|^2 + \epsilon}\left(\hat{O}(\bm{k}) - \Psi(\bm{k})\right),
  \label{eq:gradient_data}
\end{equation}
where $\epsilon$ is a small constant to prevent division by zero and $\widebar{\cdot}$ denotes the complex conjugate. A detailed derivation of this is given in~\cite{leitianMultiplexedCodedIllumination2014}. Note that we unshift frequency coordinates back by $\hat{\bm{k}}_j$ when computing the gradient. 

The derivative of the total variation regularization loss is expressed as
\begin{equation}
  \frac{\partial \mathcal{L}_\text{TV}}{\partial \hat{O}} = \mathcal{F}\left\{-\nabla \cdot \left(\frac{\nabla o(\bm{r})}{|\nabla o(\bm{r})|}\right)\right\},
  \label{eq:gradient_tv_fourier}
\end{equation}
where $\nabla \cdot$ is the divergence operator~\cite{rudin1992nonlinear}. Using Wirtinger derivatives~\cite{wirtinger1927formalen}, we compute the gradient of the phase sparsity regularization loss as
\begin{equation}
  \frac{\partial \mathcal{L}_\text{phase}}{\partial \hat{O}} = \mathcal{F}\left\{i\frac{\hat{o}(\bm{s})}{2 |\hat{o}(\bm{s})|^2+\epsilon}\right\}.
  \label{eq:gradient_phase}
\end{equation}
The full update procedure for $\hat{O}$ is thus given by
\begin{equation}
        \hat{O}(\bm{f}) \gets \hat{O}(\bm{f}) - \alpha \frac{\partial^2 \mathcal{L}_\text{data}}{\partial \hat{O}^2} - \beta \frac{\partial \mathcal{L}_\text{TV}}{\partial \hat{O}} - \gamma \frac{\partial \mathcal{L}_\text{phase}}{\partial \hat{O}},
        \label{eq:update_sample}
\end{equation}
with step sizes $\alpha$, $\beta$, and $\gamma$. In our experiments, we follow~\cite{leitianMultiplexedCodedIllumination2014} and set
\begin{equation}
  \alpha = \frac{\sum_j |M(\bm{k}+\hat{\bm{k}}_j)|}{\max(|M(\bm{k})|)}.
\end{equation}

\subsubsection{Annealing k-Space Correction}

After updating $\hat{O}(\bm{k})$, we take each k-space location $\hat{\bm{k}}_j$ and search over a local neighborhood $\hat{\bm{k}}_j + \delta$ for the offset $\delta \in \mathbb{R}^2$ which best fits the observed measurements, according to 
\begin{equation}
  \underset{\delta}{\text{argmin}} \quad \left\| \left|\mathcal{F}^{-1}\{\hat{O}(k - \hat{\bm{k}}_j + \delta) M(\bm{k})\}\right|^2 - I_j(\bm{s}) \right\|^2_2.
\end{equation}
This search is done for each shift $\hat{\bm{k}}_j$ independently to refine the k-space estimates over the course of optimization. 

In practice, we employ an linear annealing strategy for this search, starting with a large search radius $\delta_\text{max}$ and gradually reducing it to a smaller radius $\delta_\text{min}$. This approach allows for coarse corrections in the early iterations and fine-tuning in later iterations, improving both convergence speed and accuracy.

\section{Experimental Results}

\subsection{k-Space Localization}
\label{sec:k_space_localization}
\networkname{} is trained on approximately one million grayscale images from ImageNet~\cite{dengImageNetLargescaleHierarchical2009} resized to 256$\times$256. We simulate complex targets using image intensity as both amplitude and phase, where amplitude is normalized to $[0, 1]$ and phase to $[0, \pi / 4]$. Dual plane measurements are generated according to~\eqref{eq:pupil_plane} and~\eqref{eq:image_plane} with an aperture radius of 25 pixels and randomly sampled wavevectors $\bm{k}_j$ that translate the target spectrum up to $\pm60$ pixels. Optimization is done using Adam~\cite{kingmaAdamMethodStochastic2014} with a learning rate of $1\times 10^{-3}$ and a batch size of 32. Training is conducted on four NVIDIA RTX A6000 GPUs over 80k iterations, taking approximately 20 hours to complete. After training, we test the network on a held-out test set of 50,000 images and compute the root mean square error (RMSE) between the predicted k-space shifts and the ground truth.

Quantitatively, \networkname{} achieves an average RMSE of 5.04 pixels across all pupil sizes---a $5.9\%$ error with respect to the maximum possible displacement of 90 pixels.
Figure~\ref{fig:sim_comparison}(b) shows that the network accurately predicts low frequencies and high frequencies along the vertical and horizontal axes. We attribute this to the sparsity of natural images in the Fourier domain; there are less features for the network to use in sparse high frequency regions. These results demonstrate that \networkname{} is able to infer frequency shifts using structure implicit in the spectra of natural images.

\subsection{Simulated Measurements}

We evaluate the performance of our method on a small dataset of satellite images captured by the National Aeronautics and Space Administration (NASA) consisting of five scenes: the Hubble Space Telescope, the SpaceX Dragon capsule, the James Webb Space Telescope, and two images of the International Space Station (ISS). For each scene, we simulate measurements using a target width of 100m, pupil radius of 10m, and an incident wavelength of 532nm. Phase shifts are simulated using Equation~\eqref{eq:phase_shift_sample_tilt_approx} with a regularly spaced $11\times 11$ grid of tilt angles up to $\pm 0.000009^\circ$, yielding approximately $54\%$ overlap between neighboring subspectra. 

Reconstruction is performed with regularization weights $\beta = 1 \times 10^{-1}, \gamma = 1 \times 10^{-3}$. The local search algorithm is initialized with search radius of $\delta_\text{max} = 9$ pixels that anneals to $\delta_\text{min} = 1$ pixels.
To reduce computational costs, we run the search algorithm once every 10 iterations. Optimization is performed over 100 iterations, taking approximately one minute per scene.

In Figure~\ref{fig:sim_comparison}, many of the Hubble Space Telescope's features---like its antennae and solar arrays---are barely distinguishable in the low resolution measurement. These features are clearly resolved in the recovered wavefront, and  target phase is recovered with high accuracy (albeit with some residual artifacts caused by ambiguity in low contrast regions with zero amplitude). Figure~\ref{fig:sim_comparison}(b) also shows that the majority of \networkname{}'s predictions converge to the true spatial frequency locations during the optimization process. Qualitative results for all other scenes are available in Supplement 1. 

RMSE between the recovered and ground truth targets are reported for each scene in Table~\ref{tab:sim_results}. We include reconstruction results assuming a perfectly calibrated setting for reference (i.e., the ground truth frequency shifts are known). From the results, we see that our calibration-free method performs similar to the perfectly calibrated result---achieving identical performance for amplitude recovery and $87\%$ performance for phase recovery across all scenes. 

Note that, in phase retrieval problems, there is an inherent ambiguity in the recovered phase due to the fact that intensity measurements are invariant to global phase shifts. For fair comparison in our quantitative and qualitative results, we account for this ambiguity by estimating the global phase offset between our recovered phase and the ground truth phase. Specifically, we calculate the mean phase error across all pixels and subtract this offset from our recovered phase before computing error metrics. This approach ensures that our evaluation focuses on the spatial variations in phase rather than arbitrary global offsets that are inherently unrecoverable from intensity measurements alone.

\begin{table}[!t]
\renewcommand{\arraystretch}{1.3}
\caption{\textbf{Quantitative results on simulated measurements.} Best scoring results are highlighted in \textbf{bold}.}
\label{tab:sim_results}
\centering
\begin{tabular}{l||c|c||c|c}
\hline
\multirow{2}{*}{Scene} & \multicolumn{2}{c||}{Amplitude RMSE ($\downarrow$)} & \multicolumn{2}{c}{Phase RMSE ($\downarrow$)} \\
\cline{2-5}
 & ISA-FP & Calibrated & ISA-FP & Calibrated \\
\hline\hline
ISS \#1 & \textbf{0.04} & \textbf{0.04} & 1.36 & \textbf{1.23} \\
\hline
ISS \#2 & \textbf{0.04} & \textbf{0.04} & 1.35 & \textbf{1.31} \\
\hline
Webb & \textbf{0.03} & \textbf{0.03} & {1.41} & \textbf{1.29} \\
\hline
Hubble & \textbf{0.02} & \textbf{0.02} & 0.38 & \textbf{0.30} \\
\hline
Dragon & \textbf{0.03} & \textbf{0.03} & 1.54 & \textbf{1.34} \\
\hline
\end{tabular}
\end{table}
\begin{figure*}
    \centering
    \includegraphics[width=\textwidth]{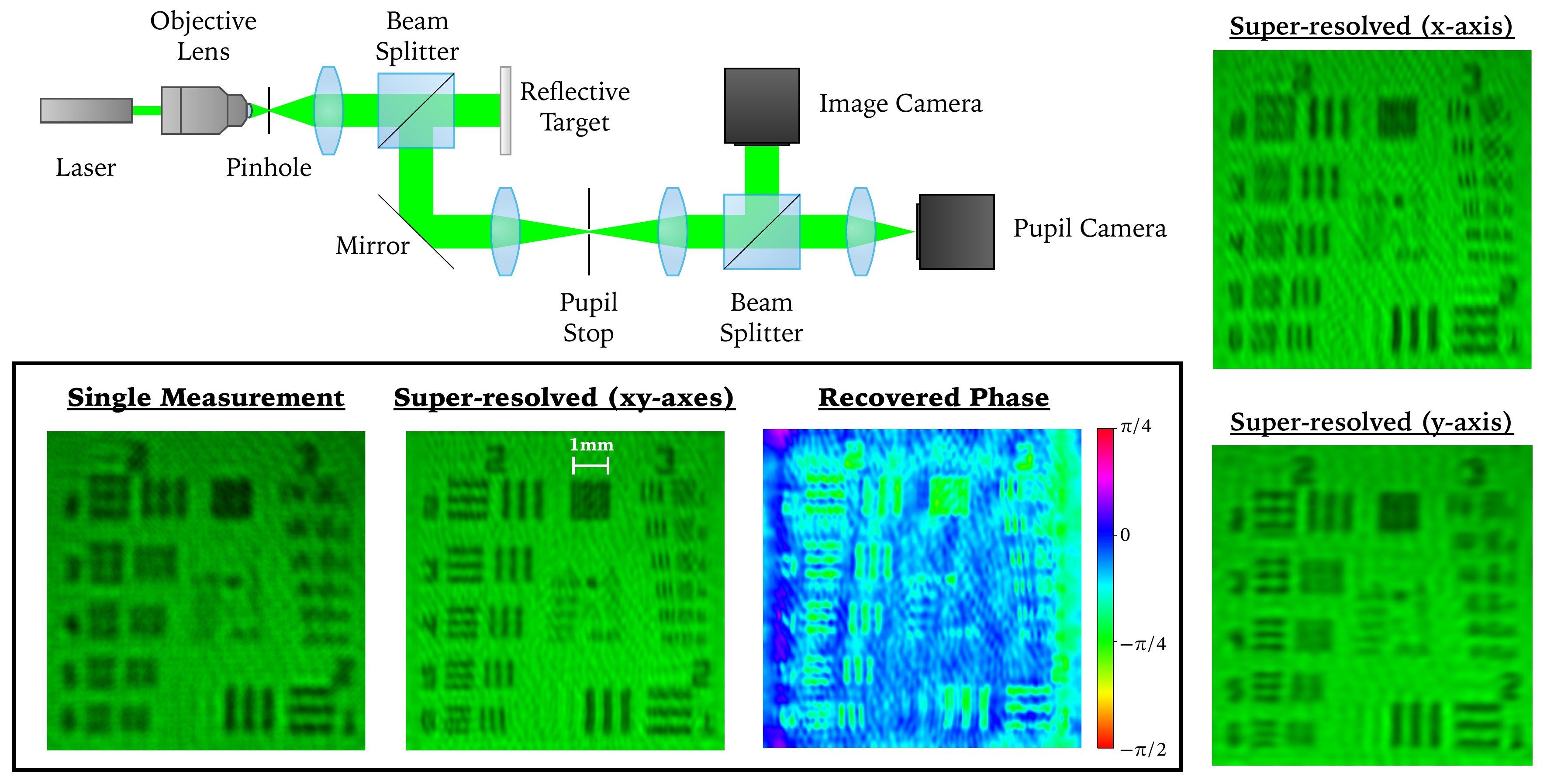}
    \caption{\textbf{Lab optical setup and experimental results.} (a) Diagram of our lab experiment imaging system. (b) We obtain dual plane measurements by scanning along the $x$ and $y$ spatial frequency axes. Applying our method to measurements captured along the $x$-axis, $y$-axis, and both axes, we super-resolve a resolution target and recover its phase.}
    \label{fig:lab_result}
\end{figure*}

\subsection{Lab Measurements}

We construct a lab optical system consisting of a positive reflective resolution test target (Thorlabs R3L1S4PR) and two cameras (FLIR BFS-U3-120S4C-CS), one at the pupil plane and the other at the image plane. The target is placed on a piezo-electric driven kinematic mount with up to $\pm 4^\circ$ tip / tilt and illuminated by a 532nm coherent laser diode (Thorlabs CPS532). The reflected wavefront is propagated through a 4f system consisting of a $f=400$mm lens, a $f=200$mm lens, and a $2$mm diameter pupil stop (for an effective $f$-stop of $f/100$). A beam splitter is used to project the wavefront onto both cameras, and an additional $f=75$mm lens is used to image the pupil plane. The diffraction limit for this system is 
\begin{align*}
  \frac{1.22\lambda}{n \times NA} &= \frac{1.22 \times 532\text{nm}}{1 \times \sin(\arctan((2\text{mm}/2)(400\text{mm})))} \\
  &\approx 0.25\text{mm}
\end{align*}

We crop measurements to a $1300\times 1300$ square and rescale them to $256 \times 256$ resolution, for an effective pixel pitch of 9.4$\mu$m. After scanning horizontal and vertical frequencies to collect 11 measurements, we recover the target wavefront with regularization weights $\beta = 1 \times 10^{-1}, \gamma = 1 \times 10^{-3}$ and an annealing schedule of $\delta_\text{max} = \delta_\text{min} = 1$. Optimization is performed over 100 iterations, taking approximately 10 seconds.

Reconstruction results are shown in Figure~\ref{fig:lab_result}. In the low resolution image, the maximum resolvable feature in the resolution chart is ${\approx}0.25$mm (Group 2 Element 1), which matches our system's theoretical diffraction limit. After reconstruction, the maximum resolvable feature increases to ${\approx}0.07$mm (Group 3 Element 6)---nearly a $4\times$ super-resolution improvement in the $x$-direction and a $2\times$ improvement in the $y$-direction. \networkname{} predictions are more accurate along the $x$-direction, a result which we attribute to optical aberrations that are not explicitly modelled in our forward model. 
In the recovered phase image, we observe the bar patterns as a phase contrast against the background. This is expected as our target is composed of low reflectance chrome bars on a higher reflectance chrome substrate.

To show the impact of better k-space initialization, we also reconstruct the target with manually calibrated k-space locations (see Supplement 1). This result achieves approximately $4\times$ super-resolution in both directions.

\section{Ablation Studies}
\label{sec:ablation}

To study the importance of input features on \networkname{}'s performance, we train our network separately on: (\textit{i}) image plane intensity, (\textit{ii}) pupil plane auto-correlation intensity, and (\textit{iii}) pupil plane intensity. Each instance of \networkname{} is trained and evaluated on a small dataset using the same procedure as in Section~\ref{sec:k_space_localization}. 

\begin{figure}[!t]
  \centering
  \includegraphics[width=0.7\columnwidth]{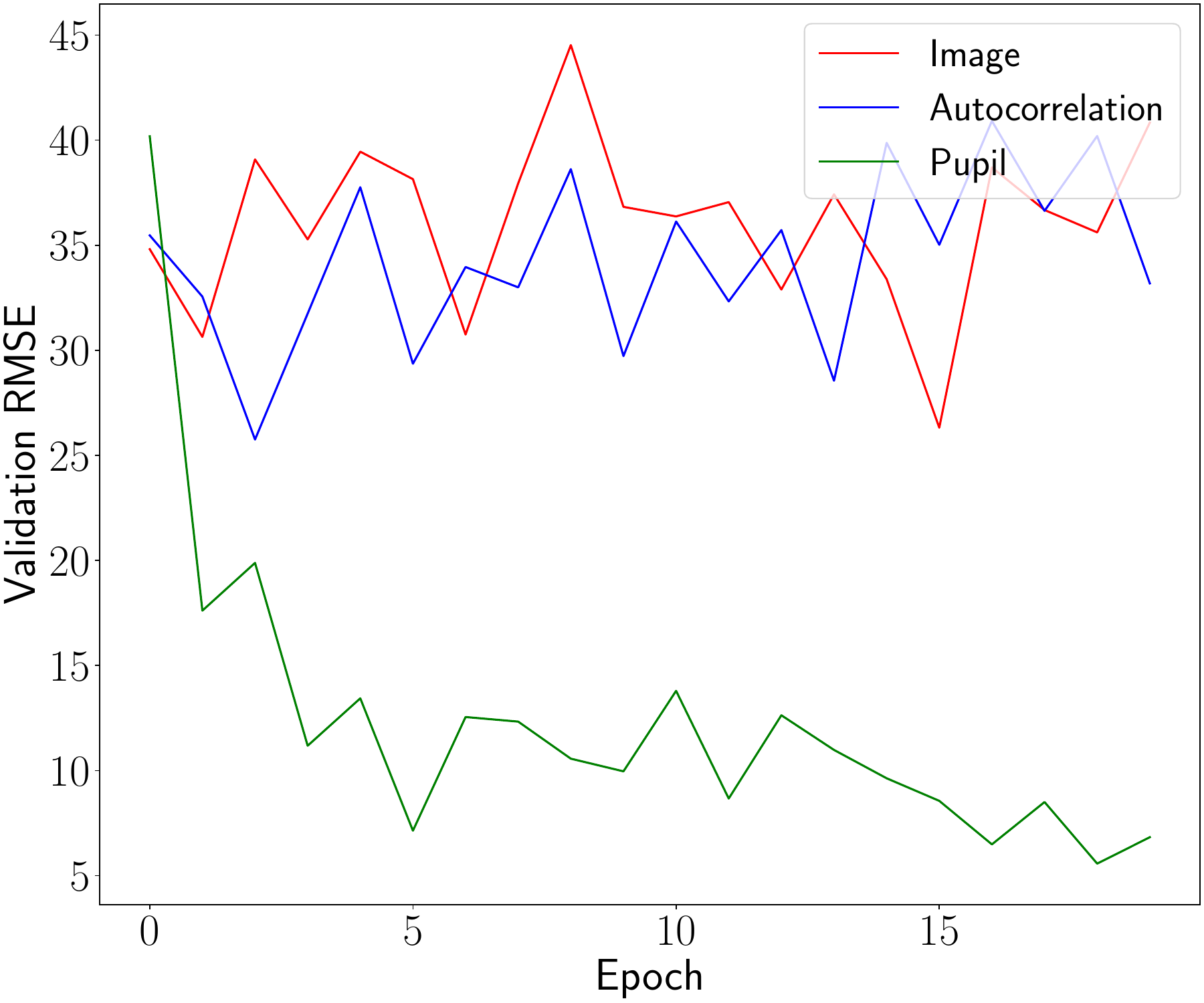}
  \caption{\textbf{Effect of input features on network learning.} \networkname{} is trained on three separate inputs: image plane intensity, auto-correlation intensity, and pupil plane intensity. Only the network trained on pupil plane intensity converges.}
  \label{fig:ablation}
\end{figure}

Figure~\ref{fig:ablation} shows the training and validation curves for each training configuration. Results clearly show that both image plane and auto-correlation features contain insufficient information to learn a generalized prior for localization. Test time RMSE for \networkname{} predictions are 34.29 and 34.96 pixels when trained on image plane and auto-correlation data respectively. Training on pupil plane intensity images reduces the test error to 10.04 pixels, suggesting that pupil plane features provide the most useful information for accurate k-space localization. 

\section{Discussion and Limitations}

While our work shows promising results, it has four significant limitations. 
First, in order to densely sample all of k-space, our approach requires that the sample rotates along both the $x$ and $y$ axis. This inhibits reconstruction of dynamic scenes using methods such as space-time FP~\cite{sun2024space} since rapid rotation introduces motion blur during acquisition. Moreover, for samples that rotate along only a single axis, we anticipate our approach will only super-resolve along one axis, as shown in Figure~\ref{fig:lab_result}.

Second, we assume little to no phase aberrations during the measurement process. However, such aberrations (e.g., atmospheric turbulence~\cite{chanComputationalImagingAtmospheric2023}) can significantly impact image quality in astronomical imaging applications. Future work should focus on incorporating aberration correction techniques~\cite{fengNeuWSNeuralWavefront2023} to make the method more robust for real-world applications. 

Third, FP is fundamentally designed for thin samples, while astronomical objects like satellites have 3D structure. 
Explicitly modeling the 3D structure of these objects using tomographic techniques~\cite{horstmeyerDiffractionTomographyFourier2016} could potentially yield better reconstructions, especially for complex spacecraft with significant depth variations. 

Last, we assume that the imaged sample is reflective. The optical path length will not change for rotating transmissive samples, and our approach is not expected to work in transmission-mode.

\section{Related Work}

\subsection{Calibration}

Calibration of optical system parameters are critical for accurate sample estimation in FP. Many works address the problem of accurate reconstruction with imprecise calibration of illumination angles~\cite{sunEfficientPositionalMisalignment2016,zhouFastRobustMisalignment2018,biancoMiscalibrationTolerantFourierPtychography2021,eckertEfficientIlluminationAngle2018}, the pupil function~\cite{ouEmbeddedPupilFunction2014}, sample motion~\cite{bianMotioncorrectedFourierPtychography2016}, and etc.~\cite{youSelfcalibratingFourierPtychographic2025,chenUncertaintyAwareFourierPtychography2025}. Nonetheless, all prior works require some form of calibration to initialize the spatial frequency shifts, which is a key advantage of our method, as it fully eliminates the need for such calibration.

\subsection{Synthetic Aperture Imaging}

Synthetic aperture (SA) imaging~\cite{cumming2005digital,pellizzariSyntheticAperatureLADAR2017} is a technique for high-resolution imaging by synthesizing a large aperture from a series of smaller measurements. Unlike ISA imaging, SA does not require target motion. Instead, the aperture is synthesized by scanning the sensor across a static target (e.g., an aircraft scanning the ground). Prior FP methods fall under the category of SA imaging, as they use moving illumination~\cite{zhengWidefieldHighresolutionFourier2013,xiangCoherentSyntheticAperture2021} or moving sensors~\cite{dongAperturescanningFourierPtychography2014,hollowaySAVISyntheticApertures2017,liFarFieldSyntheticAperture2023, songPtychoendoscopyLenslessUltrathin2024} to synthesize a large aperture. However, our method shifts this paradigm and combines FP with ISA imaging to synthesize a large aperture using target motion, which has key advantages for certain applications like astronomical imaging.

\subsection{Reflective Fourier Ptychography}

Due to its roots in microscopy, most FP techniques focus on the imaging of biological samples, which are often transparent. However, some prior works~\cite{pachecoReflectiveFourierPtychography2016,leeReflectiveFourierPtychographic2019,Ahn:21} have proposed reflection mode FP to enable imaging of opaque samples (e.g., metals, semiconductors, plastics) for surface inspection. These
methods require complex optics (e.g., parabolic mirrors) to direct illumination onto the sample at different angles. Compared to these methods, our approach is much simpler to construct in practice and does not require complex optical elements.

\subsection{Machine Learning for Fourier Ptychography}

Several deep learning methods have been applied to FP to improve reconstruction quality and reduce computational costs. 
PtychoNN~\cite{cherukaraAIenabledHighresolutionScanning2020,duPredictingPtychographyProbe2024} predicts sample amplitude and phase from ptychographic measurements. Iterative phase retrieval is performed on half of a sample scan to build the training dataset, and the network is used for quick inference on the remaining half.
Deep spatiotemporal prior (DSTP)~\cite{bohra2023dynamic} uses a deep image prior to both spatially and temporally regularize dynamic FP reconstructions.
Wang et al.~\cite{wang2022snapshot} use a U-Net to reconstruct high-resolution images from undersampled snapshot Fourier measurements.
FPM-INR~\cite{zhouFourierPtychographicMicroscopy2023,chanSparseColorFourier2024,metzlerSyntheticApertureImaging2024} leverages implicit neural representations (INRs) to efficiently represent and reconstruct a target scene. Our method differs from existing learning-based FP methods in that it does not directly reconstruct the target, but instead maps measurements to their corresponding k-space locations, reducing the need for precise calibration.

\section{Conclusion}

In this work, we introduced inverse synthetic aperture Fourier ptychography (ISA-FP), a novel approach that enables high-resolution imaging by leveraging target motion rather than illumination angle diversity. ISA-FP significantly broadens the applicability of FP to scenarios where illumination cannot be directly controlled, such as astronomical imaging. By exploiting the natural rotation of targets, our method achieves comparable reconstruction quality to traditional calibrated approaches while eliminating the need for precise illumination angle control.

We also introduced \networkname{}, a neural network that estimates spatial frequency shifts from dual plane intensity measurements. This network demonstrates strong generalization capabilities to natural images, achieving an average RMSE of 5 pixels. The network's ability to learn structural priors in image spectra enables calibration-free FP, making the technique more robust and practical for real-world applications. Our experimental validation demonstrates that ISA-FP can achieve $3.6\times$ super-resolution compared to conventional imaging methods.

The impact of our work extends beyond the specific application of astronomical imaging. By removing the requirement for precise calibration and enabling passive illumination schemes, ISA-FP opens new possibilities for high-resolution imaging in diverse fields where traditional FP would be impractical. 
Future work could explore incorporating aberration correction techniques to further enhance the method's robustness. Extending our method for passive super-resolution (i.e., using ambient illumination) is another promising avenue of research. We hope our implementation and hardware designs will facilitate broader adoption and continued development of this promising imaging technique.

\begin{backmatter}
\bmsection{Funding} 
Office of Naval Research (ONR) (N00014-23-1-2752); Air Force Office of Scientific Research (AFOSR) Young Investigator Program (FA9550-22-1-0208); Joint Directed Energy Transition Office (JDETO).

\bmsection{Acknowledgment} 
M.A.C. thanks Janith~B.~Senanayaka, University of Maryland, for insightful discussions on Fourier optics and Fourier ptychography.
M.A.C. and C.A.M. were supported in part by AFSOR, ONR, and JDETO.

\bmsection{Disclosures} The views expressed in this article are those of the authors and do not necessarily reflect the official policy or position of the United States Air Force Academy, the Air Force, the Department of Defense, or the U.S. Government. This work is approved for public release; distribution is unlimited. Public Affairs release approval \#USAFA-DF-2025-471.

\bmsection{Disclosures} The authors declare no conflicts of interest.

\bmsection{Supplemental document}
See Supplement 1 for additional results.

\end{backmatter}

\bibliography{main}

\clearpage

\setcounter{figure}{0}
\renewcommand{\thefigure}{A\arabic{figure}}
\setcounter{table}{0}
\renewcommand{\thetable}{A\arabic{table}}
\setcounter{equation}{0}
\renewcommand{\theequation}{A\arabic{equation}}

\appendix

\section{Converting Rotation to Pixel Translation}

A fundamental property of the Fourier transform is that a linear phase ramp applied in the spatial domain corresponds to a translation in the frequency domain. This relationship is critical for understanding how sample rotation in our experimental setup translate to shifts in the measured spectrum.

Consider a wavefront $o(\bm{s})$ modulated by a spatially dependent phase ramp:
\begin{equation}
  o'(\bm{s}) = o(\bm{s}) \cdot \exp\left\{ i \bm{s} \cdot \frac{2 \pi}{\lambda} \bm{\phi} \right\},
\end{equation}
where $\bm{s}$ is the spatial coordinate vector, $\lambda$ is the wavelength, and $\bm{\phi}$ represents the phase gradient. According to the Fourier shift theorem, the Fourier transform of this modulated wavefront is:
\begin{equation}
  \hat{O}'(\bm{k}) = \hat{O}(\bm{k} - \bm{k}_j),
\end{equation}
where $\hat{O}$ is the Fourier transform of the original wavefront and the frequency shift $\bm{k}_j$ is given by:
\begin{equation}
  \bm{k}_j = \frac{\bm{\phi}}{\lambda},
\end{equation}
with units of cycles per meter (spatial frequency).

In digital implementations, we need to convert this physical frequency shift to pixel shifts in the discretely sampled spectrum. If our spatial domain is sampled with $N \times N$ pixels at a pixel pitch of $\Delta x$, then the corresponding frequency domain has a sampling pitch of
\begin{equation}
  \Delta f = \frac{1}{N \Delta x},
\end{equation}
which represents the smallest resolvable frequency difference between adjacent pixels in the spectrum.
The number of pixels that the spectrum shifts in the discrete Fourier domain can therefore be calculated as
\begin{equation}
  \frac{\bm{k}_j}{\Delta f} = \frac{\bm{\phi}}{\lambda} \cdot N \Delta x.
\end{equation}
This relationship allows us to precisely map the physical tilt of our sample to the corresponding pixel shift in the captured Fourier spectrum.

Applying this mapping to \eqref{eq:phase_shift_sample_tilt_approx}, we can convert the sample rotation angle $\theta$ to the corresponding pixel shift in the frequency domain:
\begin{equation}
  \frac{\bm{k}_j}{\Delta f} = \frac{2\theta}{\lambda} N \Delta x.
\end{equation}
Similarly, we can convert a pixel shift in the frequency domain to a sample rotation angle using
\begin{equation}
  \theta = \frac{\lambda \bm{k}_j}{2 N \Delta x \Delta f}.
\end{equation}

\section{Additional Simulation Results}

We include additional qualitative simulation results for the ISS, Webb, and Dragon scenes that further validate our approach. These supplementary results demonstrate the robustness of our technique across various scenarios and highlight the effectiveness of our learning-based method for estimating k-space coordinates.

\begin{figure*}
    \centering
    \begin{tikzpicture}
        \node[inner sep=0pt] () at (-0.6,0) {\includegraphics[width=.7\textwidth]{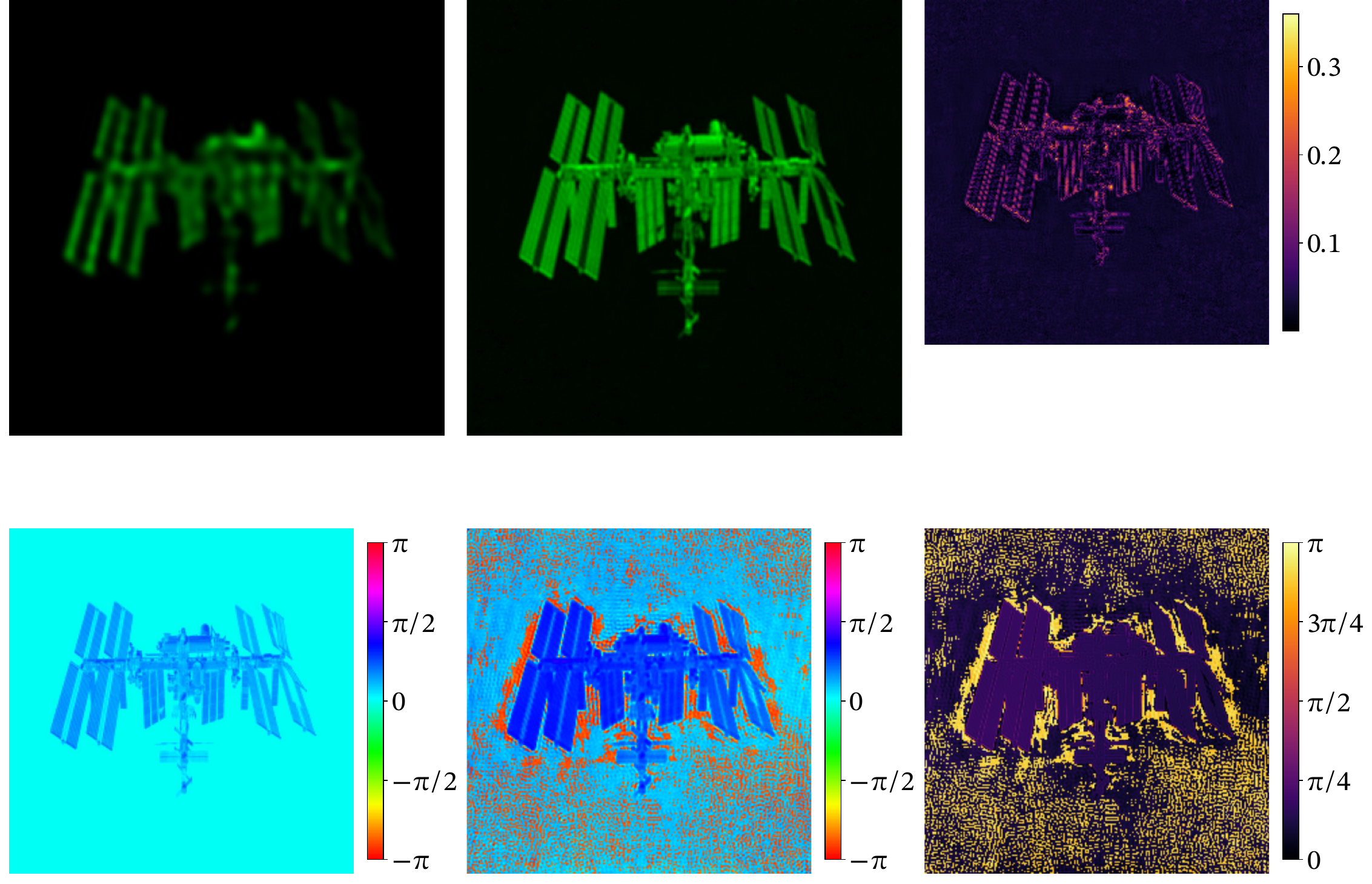}};
        \node[inner sep=0pt] () at (8.6,0) {\includegraphics[width=.21\textwidth]{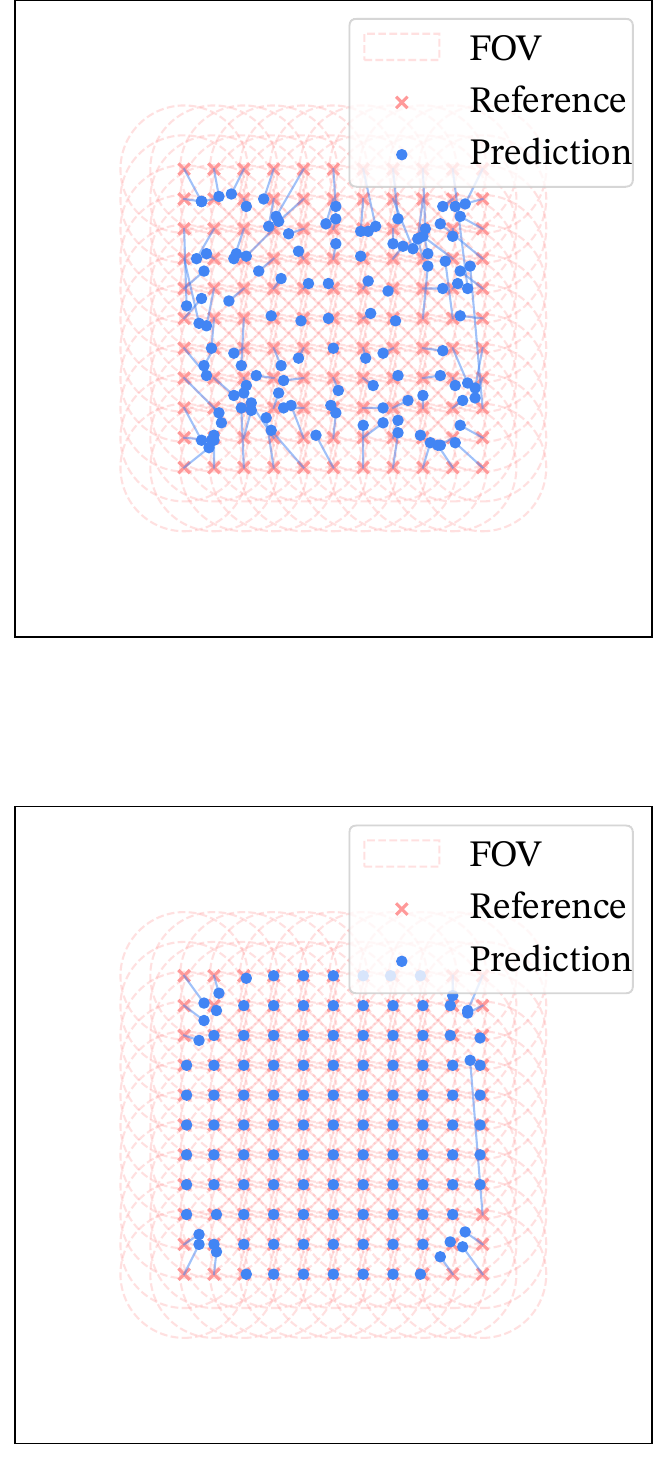}};
        \draw[thick, white, |-|] (-2.3,0.5) -- (-1.3,0.5) node[midway,above] {25m};

        \node[above] at (-4.9, 4.2) {Low Resolution Image\strut};
        \node[above] at (-0.5, 4.2) {Super-resolved\strut};
        \node[above] at (3.3, 4.2) {Amplitude Error\strut};

        \node[above] at (-5.3, -0.7) {Ground-truth Phase\strut};
        \node[above] at (-1.0, -0.7) {Recovered Phase\strut};
        \node[above] at (3.4, -0.7) {Phase Error\strut};

        \node[above] at (8.7, 4.2) {\networkname{} Output};
        \node[above] at (8.7, -0.4) {Corrected k-Space};
    \end{tikzpicture}
    \caption{\textbf{International Space Station \#1.}}
\end{figure*}

\begin{figure*}
    \centering
    \begin{tikzpicture}
        \node[inner sep=0pt] () at (-0.6,0) {\includegraphics[width=.7\textwidth]{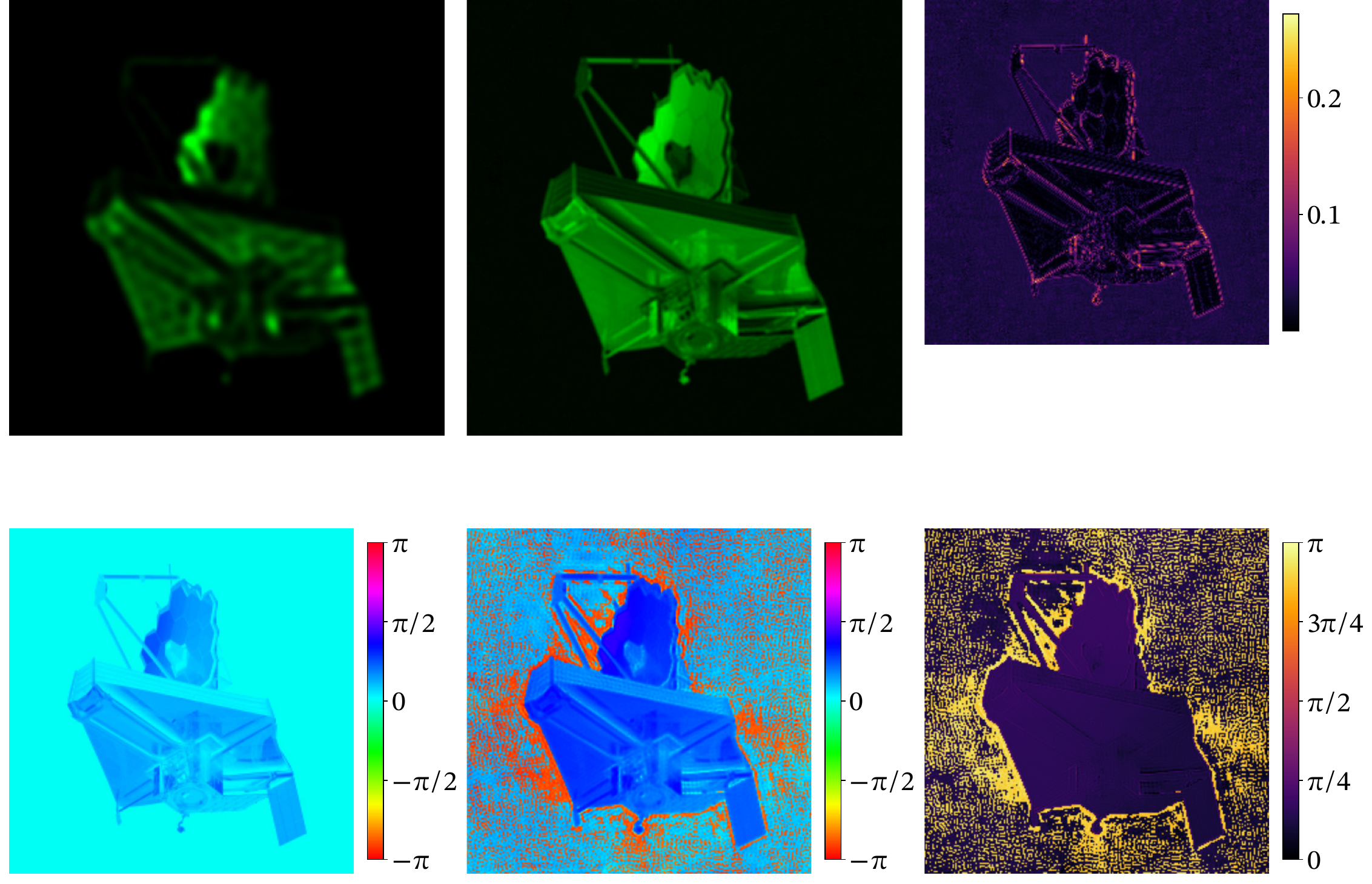}};
        \node[inner sep=0pt] () at (8.6,0) {\includegraphics[width=.21\textwidth]{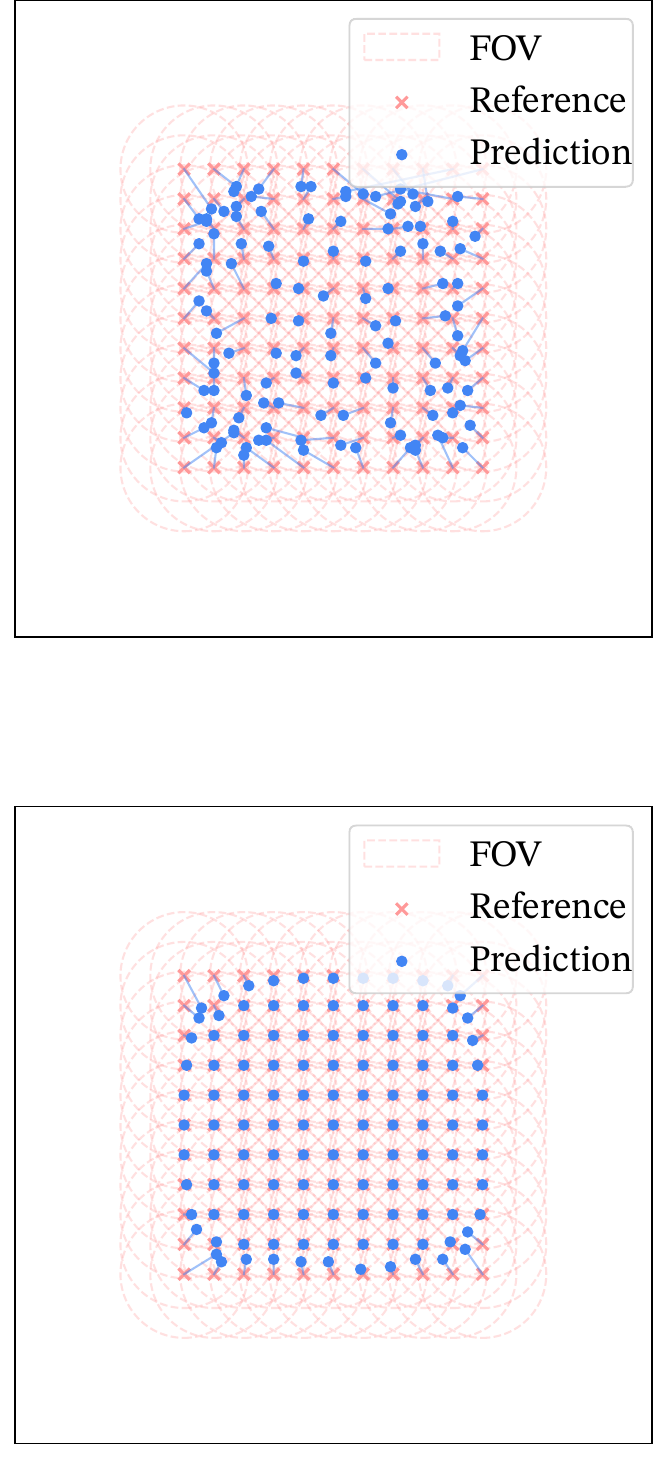}};
        \draw[thick, white, |-|] (-2.3,0.5) -- (-1.3,0.5) node[midway,above] {25m};

        \node[above] at (-4.9, 4.2) {Low Resolution Image\strut};
        \node[above] at (-0.5, 4.2) {Super-resolved\strut};
        \node[above] at (3.3, 4.2) {Amplitude Error\strut};

        \node[above] at (-5.3, -0.7) {Ground-truth Phase\strut};
        \node[above] at (-1.0, -0.7) {Recovered Phase\strut};
        \node[above] at (3.4, -0.7) {Phase Error\strut};

        \node[above] at (8.7, 4.2) {\networkname{} Output};
        \node[above] at (8.7, -0.4) {Corrected k-Space};
    \end{tikzpicture}
    \caption{\textbf{James Webb Space Telescope.}}
\end{figure*}

\begin{figure*}
    \centering
    \begin{tikzpicture}
        \node[inner sep=0pt] () at (-0.6,0) {\includegraphics[width=.7\textwidth]{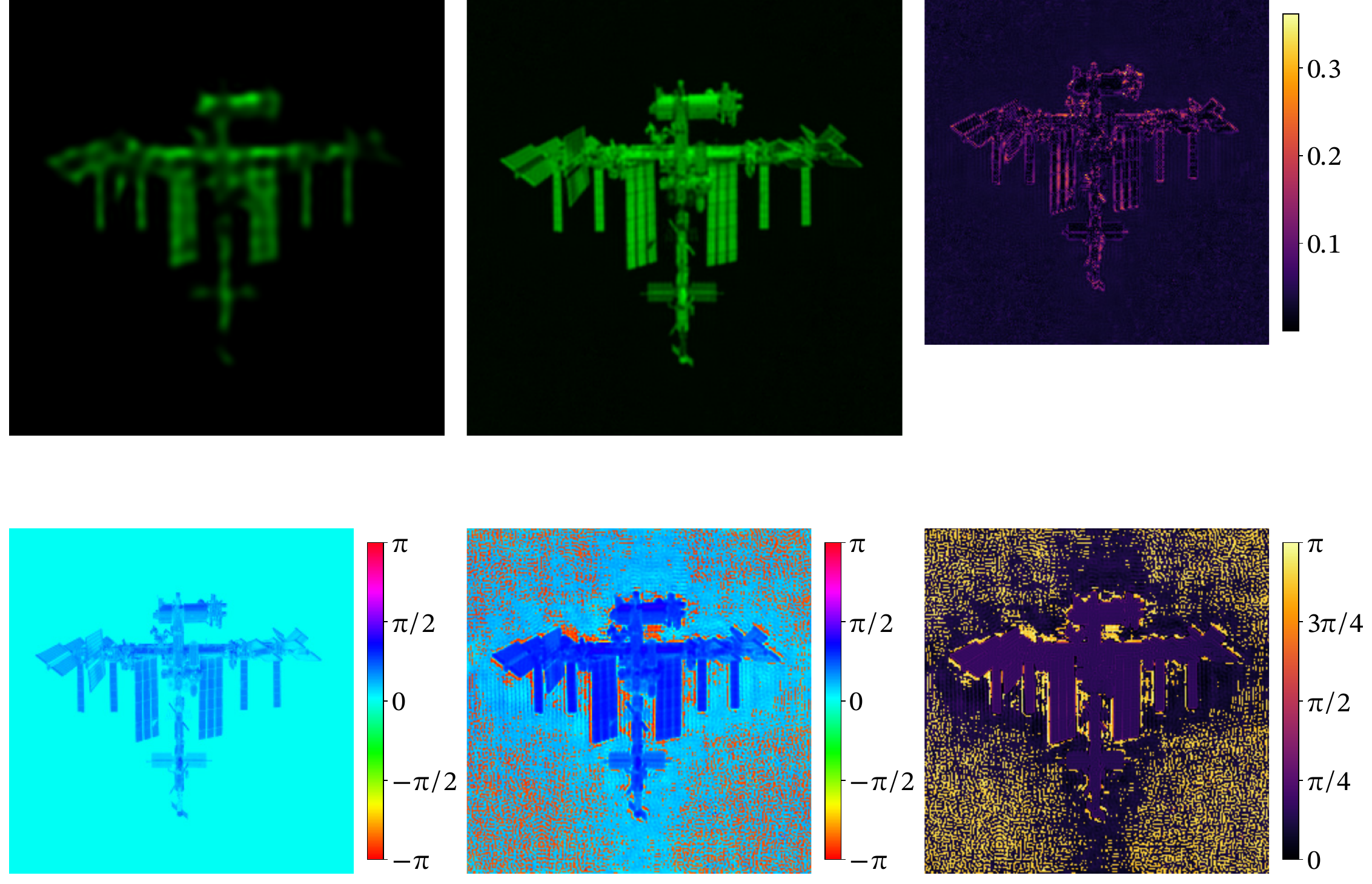}};
        \node[inner sep=0pt] () at (8.6,0) {\includegraphics[width=.21\textwidth]{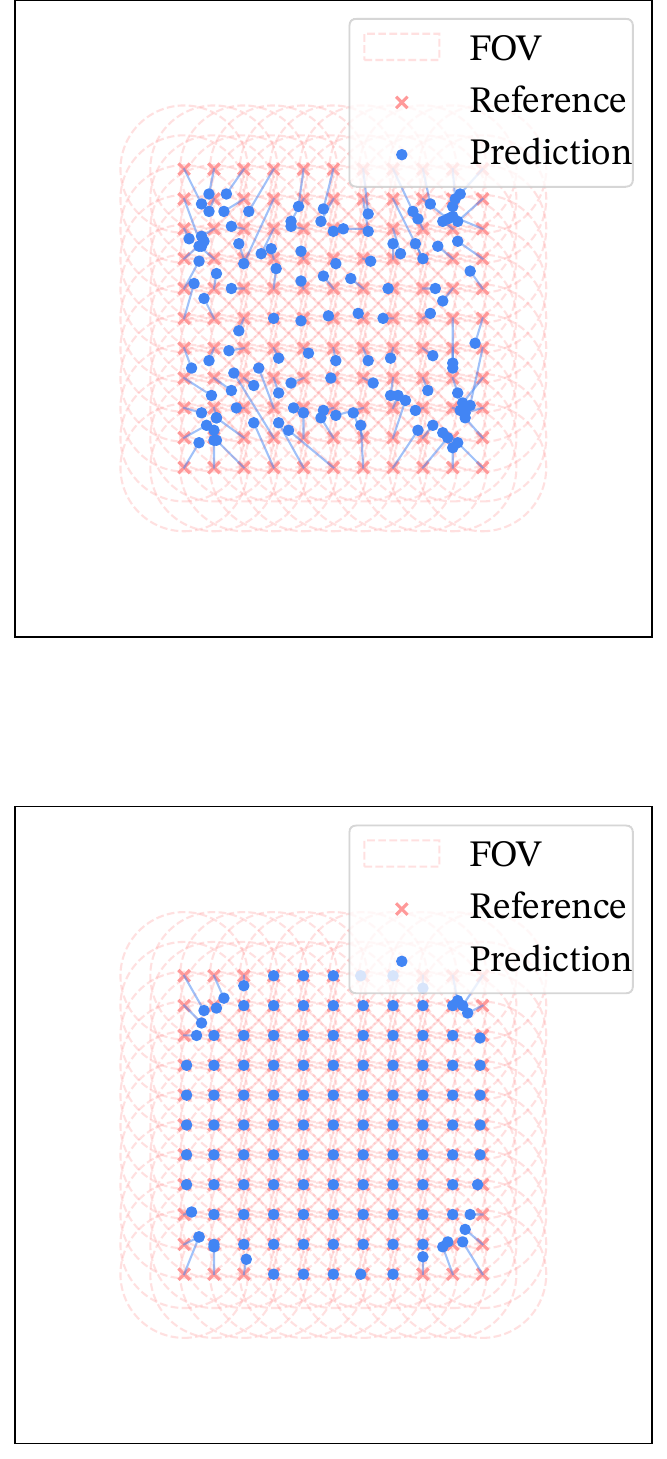}};
        \draw[thick, white, |-|] (-2.3,0.5) -- (-1.3,0.5) node[midway,above] {25m};

        \node[above] at (-4.9, 4.2) {Low Resolution Image\strut};
        \node[above] at (-0.5, 4.2) {Super-resolved\strut};
        \node[above] at (3.3, 4.2) {Amplitude Error\strut};

        \node[above] at (-5.3, -0.7) {Ground-truth Phase\strut};
        \node[above] at (-1.0, -0.7) {Recovered Phase\strut};
        \node[above] at (3.4, -0.7) {Phase Error\strut};

        \node[above] at (8.7, 4.2) {\networkname{} Output};
        \node[above] at (8.7, -0.4) {Corrected k-Space};
    \end{tikzpicture}
    \caption{\textbf{International Space Station \#2.}}
\end{figure*}

\begin{figure*}
    \centering
    \begin{tikzpicture}
        \node[inner sep=0pt] () at (-0.6,0) {\includegraphics[width=.7\textwidth]{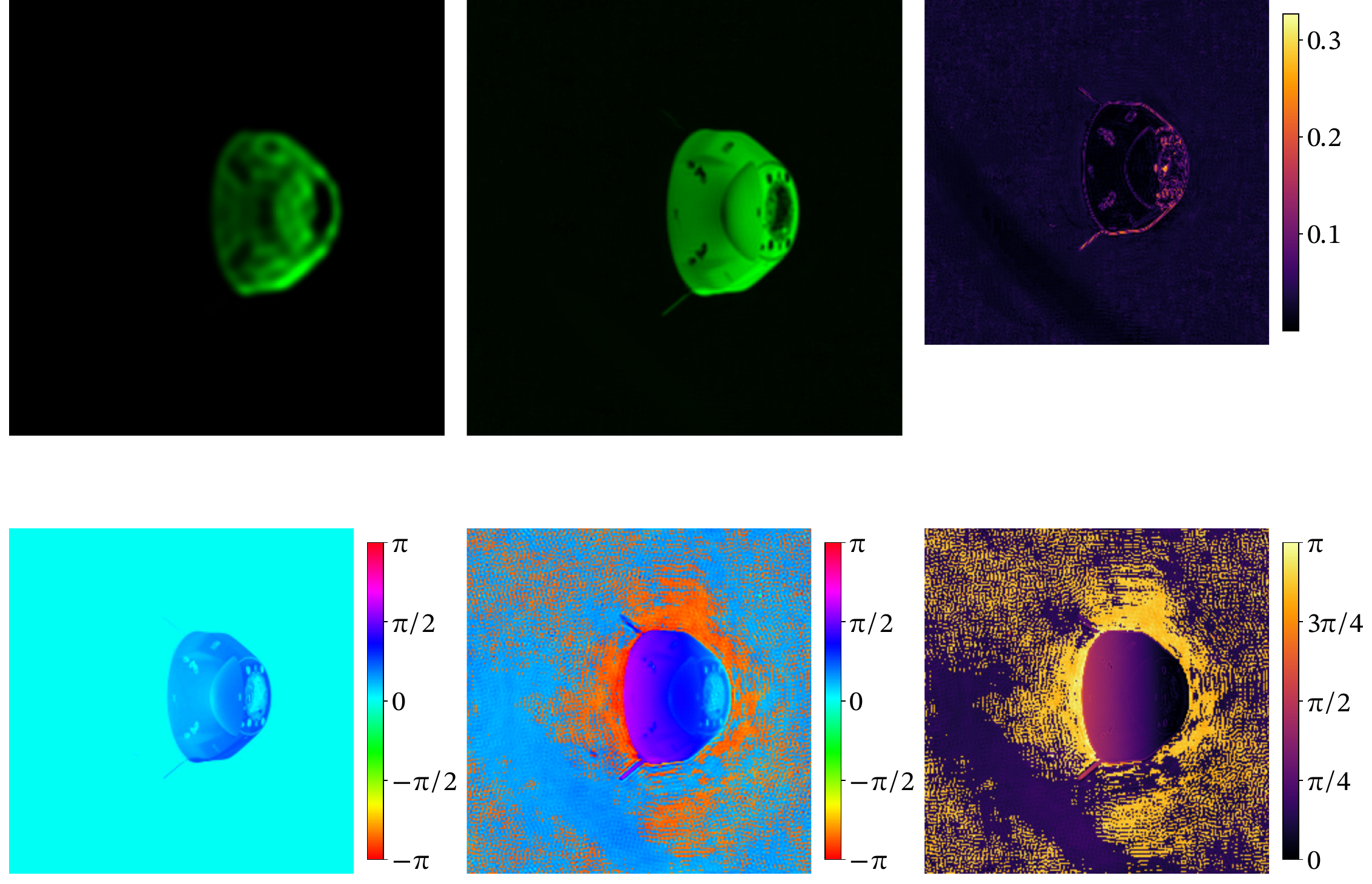}};
        \node[inner sep=0pt] () at (8.6,0) {\includegraphics[width=.21\textwidth]{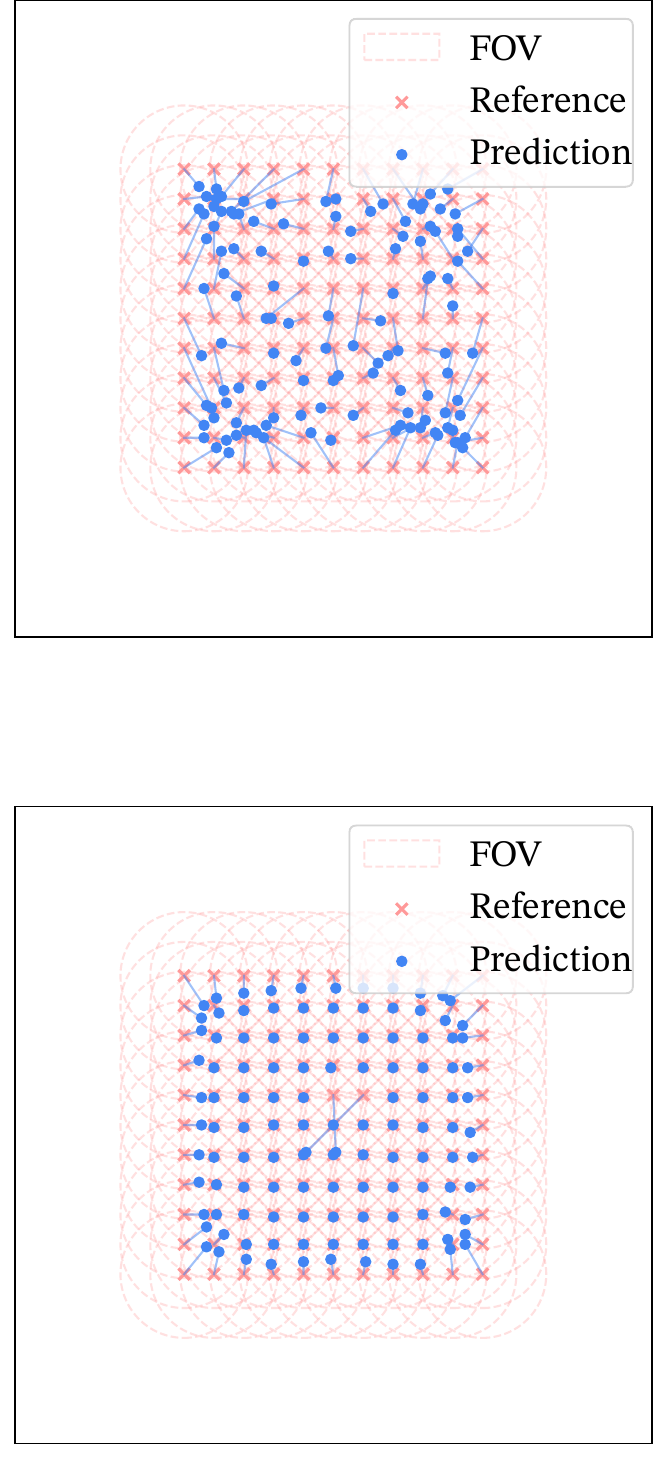}};
        \draw[thick, white, |-|] (-2.3,0.5) -- (-1.3,0.5) node[midway,above] {25m};

        \node[above] at (-4.9, 4.2) {Low Resolution Image\strut};
        \node[above] at (-0.5, 4.2) {Super-resolved\strut};
        \node[above] at (3.3, 4.2) {Amplitude Error\strut};

        \node[above] at (-5.3, -0.7) {Ground-truth Phase\strut};
        \node[above] at (-1.0, -0.7) {Recovered Phase\strut};
        \node[above] at (3.4, -0.7) {Phase Error\strut};

        \node[above] at (8.7, 4.2) {\networkname{} Output};
        \node[above] at (8.7, -0.4) {Corrected k-Space};
    \end{tikzpicture}
    \caption{\textbf{SpaceX Dragon capsule.}}
\end{figure*}

\begin{figure*}
  \centering
  \includegraphics[width=0.8\textwidth]{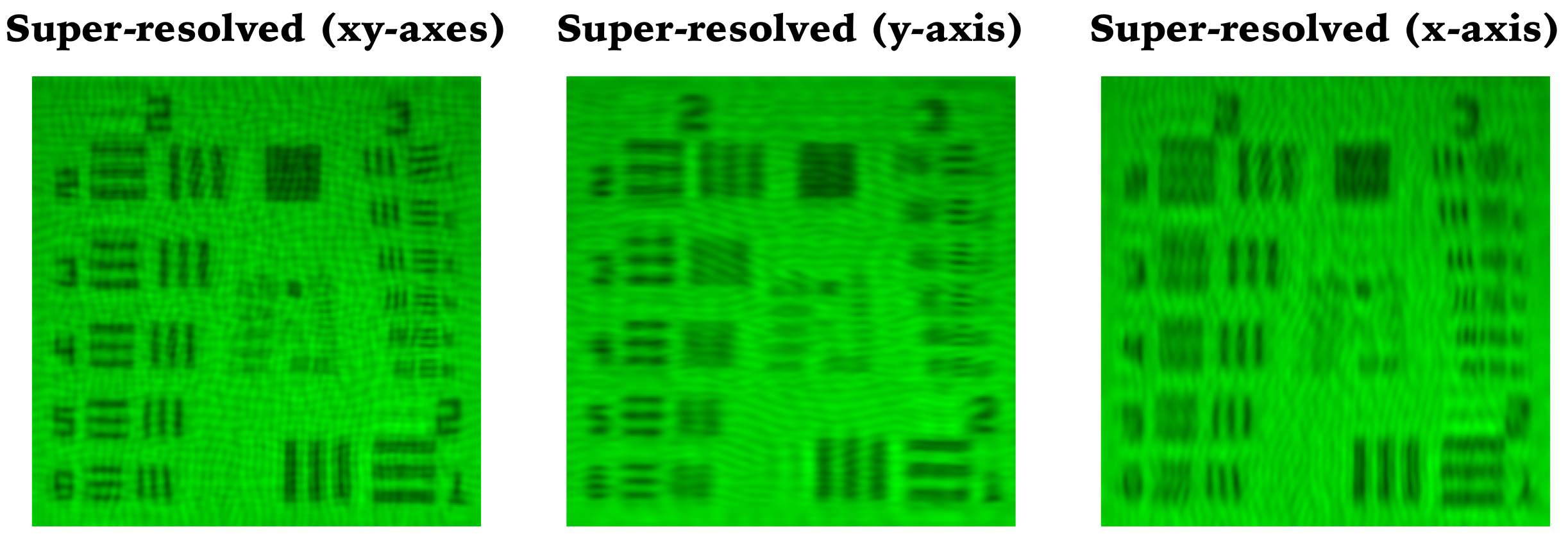}
  \caption{\textbf{Manually calibrated lab result.} For reference, we manually initialize k-space locations and reconstruct the target wavefront.}
\end{figure*}

\clearpage

\bibliographyfullrefs{main}

\end{document}